\documentclass[12pt,a4paper]{article}

\usepackage[T1]{fontenc}                   

\usepackage[utf8]{inputenc}                

\usepackage{microtype}                     

\usepackage[english]{babel} 

\newcommand{\subtitle}[1]{%
	\posttitle{%
		\par\end{center}
	\begin{center}\large#1\end{center}
	\vskip0.5em}%
}

\usepackage{hyperref}                      

\usepackage{bookmark}                      

\usepackage{url}
\usepackage[T1]{fontenc}
\usepackage{lmodern}
\usepackage{amsmath, amsthm, amssymb, amsfonts}
\usepackage{color}
\usepackage{graphicx}
\usepackage{graphics}
\usepackage{xcolor}
\usepackage{bm}
\usepackage{listings}
\usepackage{pstricks}
\def\beq{\begin{equation}}
\def\eq{\end{equation}}
\def\bea{\begin{eqnarray}}
\def\eea{\end{eqnarray}}
\newcommand{\bkappa}{\mbox{\boldmath $\kappa$}}

\newcommand{\bk}{\mbox{\boldmath $k$}}

\newcommand{\be}{\mbox{\boldmath $e$}}

\newcommand{\bI}{\mbox{\boldmath $I$}}
\newcommand{\ket}[1]{| {#1} \rangle}

\def\Pom{{\bf I\!P}}
\addto\captionsitalian{}

\usepackage[a4paper, top=2cm, left=3cm, right=3cm, left=3cm, bottom=3cm]{geometry}

\begin{document}
\begin{titlepage}
\begin{center}
\Large \textbf{Exclusive production of $\phi$ meson in the $\gamma^*\,p \to \phi\,p$ reaction at large photon virtualities\\ within $k_T$-factorization approach}
\end{center}

\vskip 0.5cm

\centerline{
	A.D.~Bolognino$^{1,2,*}$, A.~Szczurek$^{3,4,\S}$, W.~Sch\"afer$^{4,\ddagger}$}

\vskip .6cm

\centerline{${}^1$ {\sl Dipartimento di Fisica, Universit\`a della Calabria}}
\centerline{\sl I-87036 Arcavacata di Rende, Cosenza, Italy}
\vskip .2cm
\centerline{${}^2$ {\sl Istituto Nazionale di Fisica Nucleare, Gruppo collegato
		di Cosenza}}
\centerline{\sl I-87036 Arcavacata di Rende, Cosenza, Italy}
\vskip .2cm
\centerline{${}^3$ {\sl College
		of Natural Sciences, Institute of Physics,
		University of Rzesz\'ow}}
\centerline{\sl ul. Pigonia 1, PL-35-310 Rzesz\'ow, Poland}
\vskip .2cm
\centerline{${}^4$ {\sl Institute of Nuclear Physics Polish Academy of Sciences}}
\centerline{\sl ul. Radzikowskiego 152, PL-32-342, Krak\'ow, Poland}

\begin{abstract}
We apply the $k_T$-factorization approach to the production of
$\phi$ meson in deep-inelastic scattering. The helicity-conserving 
$\gamma^*(T,L) \to \phi$
impact factor is calculated for longitudinal and transverse photon polarization 
using $\phi$ meson distribution amplitudes.
Different unintegrated gluon distributions are used in the calculations.
The formalism for massless quarks/antiquarks
gives too large transverse and longitudinal cross sections for
photon virtualities below $Q^2\sim 8 \, \rm{GeV}^2$. 
We suggest how to improve the description
of the HERA data by introducing effective strange
quark masses into the formalism. We derive the corresponding 
massive impact factor by comparing to the light-cone wave function
representation used in previous $k_T$-factorization calculations 
and the color-dipole approaches. As a byproduct we present 
expressions for higher twist-amplitudes as weighted integrals
over the light-cone wave function.
The quark mass $m_q \approx 0.5$
GeV allows to improve the description of both longitudinal and transverse cross
section down to $Q^2 \sim$ 4 GeV$^2$. 
We present also the polarized cross section ratio
$\sigma_L/\sigma_T$ and the behavior of the total cross section $\sigma_{tot} = \sigma_{L} + \sigma_{T}$ as a function of photon virtuality.
\end{abstract}

\vskip .5cm

$^{*}${\it e-mail}:
ad.bolognino@unical.it

$^{\S}${\it e-mail}:
antoni.szczurek@ifj.edu.pl

$^{\ddagger}${\it e-mail}:
wolfgang.schafer@ifj.edu.pl

\end{titlepage}

\section{Introduction}

The diffractive electroproduction of vector mesons, $\gamma^*\,p \to V p$,
has attracted much attention at the HERA collider (for a review
see e.g. Ref.~\cite{Ivanov:2004ax}) and is to be expected an important subject in future experiments e.g. at an electron-ion collider (EIC) \cite{Accardi:2012qut}.
In this work we are interested in the limit of large $\gamma^*\,p$ center-of-mass energy $W$, $s\equiv W^2\gg Q^2\gg\Lambda_{\rm QCD}^2$, 
which implies small gluon longitudinal fraction $x=(Q^2+m_V^2)/(W^2+ Q^2 - m_p^2) \sim Q^2/W^2$.
In this kinematics, the photon virtuality $Q^2$ gives a handle on the dominant size of color dipoles in the $\gamma^* \to V$ transition and thus allows to study a transition from the hard, perturbative (small dipole), to the soft,
nonperturbative (large dipole), regimes of scattering.
In momentum space, the color dipole approach has its correspondence
in the $k_T$-factorization, where the main ingredient is the unintegrated (transverse-momentum dependent) gluon distribution (UGD).
At large  photon virtualities the diffractive cross section is a
sensitive probe of the proton UGD.

The $k_T$-factorization formalism reviewed in Ref.~\cite{Ivanov:2004ax}, 
includes besides the transverse momentum of gluons also the 
transverse momentum of quark and antiquark in the vector meson as
encoded in the light-cone wave function of the meson.
This approach was used with some success in Ref. \cite{Cisek:2010jk}. At very large $Q^2$ one may expect, that the 
relative transverse motion of (anti-)quarks in the bound state becomes negligible, and can be integrated out. Then, regarding the vector meson
only a dependence on the longitudinal momentum fraction of quarks
encoded in distribution amplitudes (DA) is left. 

A quite general factorization formalism of vector meson production
in deep inelastic scattering was formulated in Refs.~\cite{Anikin:2009bf,Anikin:2011sa} and has been recently applied
in Ref.~\cite{Bolognino:2018} to the diffractive deep inelastic production
of $\rho$ mesons. While the production of longitudinal vector mesons
involves the leading twist DA, similar to the one introduced for other
hadronic processes in Refs.~\cite{Radyushkin:1977gp,Lepage:1979zb,Lepage:1980fj,Chernyak:1983ej},
for transverse vector mesons higher twists are involved and corresponding DAs studied in Ref.~\cite{Ball:1998sk} are needed.

The recent analysis of helicity amplitudes for $\rho^0$
meson production~\cite{Bolognino:2018} showed
that this approach may be useful in testing UGDs.
In this paper, in order to test the formalism further, we wish to focus
on and investigate the exclusive photoproduction of $\phi$ meson:
\begin{center}
$\gamma^* \, p \rightarrow \phi \, p$ \,.
\end{center}
Corresponding experimental data were obtained by 
the H1~\cite{Adloff:2000nx,Aaron:2009xp} and ZEUS~\cite{ZEUScoll2005} collaborations at HERA.
In this paper we will show how the $k_T$-factorization approach 
of \cite{Ivanov:2000cm} matches to the higher-twist DA expansion,
at least in the Wandzura-Wilczek (WW) approximation, where no explicit 
$q \bar q g$ contributions are included.
Future applications at EIC may require the inclusion of
next-to-leading-order contributions. Here the approach based on DAs
has the advantage that for the case of collinear partons in the final state, the necessary techniques for the calculation of NLO impact factors are well advanced \cite{Ivanov:2004pp,Ivanov:2005gn}. 
The first part of the paper is devoted to the summary of the theoretical
framework of calculating the helicity amplitudes within 
the  $k_T$-factorization approach.
The second part shows the cross sections of the process and 
the effects due to UGDs and/or due to the strange-quark mass.
Here we take advantage of the fact that we can rather straightforwardly
derive the massive impact factor in the WW approximation from the 
light-cone wave function approach.
As a byproduct we show how higher twist DAs can be obtained from the
light-cone wave functions which may be interesting for the application
of various light-cone models also to other vector mesons. 

A comparison with the H1 and ZEUS measurements will be presented.
The conclusion section will close our paper.

\section{Theoretical framework}

\subsection{Helicity-amplitudes $\mathbf{T_{\lambda_V\lambda_\gamma}}$}

 In the high-energy regime, $s\equiv W^2\gg Q^2\gg\Lambda_{\rm QCD}^2$, which implies small $x=(Q^2+m_V^2)/(W^2+ Q^2 - m_p^2) \sim Q^2/W^2$, the forward helicity amplitude $T_{\lambda_V\lambda_\gamma}$ can be expressed, in the $k_T$-factorization, as
 the convolution of the $\gamma^*\rightarrow V$ impact factor (IF),
 $\Phi^{\gamma^* \to V}_{\lambda_V,\lambda_\gamma}(\bkappa^2,Q^2)$, with the UGD, ${\cal F}(x,\kappa^2)$.
 
 Our normalization of the impact factor is chosen, such that the forward amplitude 
 for the $\gamma^*\,p \to V\,p $ process reads
 \begin{equation}
 \label{amplitude}
 \Im m T_{\lambda_V\lambda_\gamma}(s,Q^2) = s \int \dfrac{d^2\bkappa}
 {(\bkappa^2)^2}\Phi^{\gamma^* \to V}_{\lambda_V,\lambda_\gamma}(\bkappa^2,Q^2)
 {\cal F}(x,\bkappa^2).\quad 
 \end{equation}
 Here, the UGD is related to the collinear 
 gluon parton distribution as
 \begin{eqnarray}
 xg(x,\mu^2) = \int^{\mu^2} {d \bkappa^2 \over \bkappa^2} {\cal F}(x,\bkappa^2) \, .
 \end{eqnarray}

We now turn to the two different approaches to the impact factors 
which we want to compare in this work.

\subsection{Distribution amplitude expansion}

We start with the scheme based on the collinear factorization of the
meson structure which was worked out in Refs.~\cite{Anikin:2009bf,Anikin:2011sa} and was used recently for $\rho$-meson electroproduction in Ref.~\cite{Bolognino:2018}. 
This approach starts from the observation, that at large $Q^2$ 
the transverse internal motion of partons in the meson can be neglected.
 
The longitudinal impact factor is expressed in terms of the 
standard twist-2 distribution amplitude.
In the normalization adopted by us, the IF for the $L \to L$ transition 
reads
 	\begin{equation}
 	\label{Phi_LL}
 	\Phi^{\gamma^* \to V}_{0,0}(\bkappa^2,Q^2) = {4 \pi \alpha_S e_q \sqrt{4 \pi \alpha_{\rm em}} f_V \over N_c Q}
 	\int^{1}_{0}dy\, \varphi_1(y;\mu^2)\left(\frac{\alpha}{\alpha + y\bar{y}}\right)
 	\,,
 	\end{equation}
 	where $\alpha = \bkappa^2/Q^2$, $y$ is the fraction of the mesons's lightcone-plus momentum 
 	carried by the quark, $\bar y = 1-y$, and $\varphi_1(y;\mu^2)$ is the twist-2 distribution amplitude (DA). It is normalized as
 	\begin{eqnarray}
 	\int_0^1 dy \, \varphi_1(y;\mu^2) = 1
 	\end{eqnarray}
 	 and we recall its asymptotic form 
 	\begin{equation}
 	\label{phi}
 	\varphi_1(y; \mu^2) \xrightarrow{\mu^2\rightarrow \infty} \varphi_1^{as}(y) = 6y\bar{y} \, .
 	\end{equation}

 The expression for the transverse case
 is:
 
 \begin{eqnarray}
 	\Phi^{\gamma^* \to V}_{+,+}(\bkappa^2,Q^2) &=& {2 \pi \alpha_S e_q \sqrt{4 \pi \alpha_{\rm em}} f_V m_V\over N_c Q^2} \nonumber \\
 	&\times& \left\{ \int^{1}_{0} dy \frac{\alpha (\alpha +2 y \bar{y})}{y\bar{y}
 		(\alpha+y\bar{y})^2}\right. \left[(y-\bar{y})\varphi_1^T(y;\mu^2)
 	+ \varphi_A^T(y;\mu^2)\right] \nonumber \\
 	&-&\int^{1}_{0}dy_2\int^{y_2}_{0}dy_1 \frac{y_1\bar{y}_1\alpha}
 	{\alpha+y_1\bar{y}_1} \nonumber \\
 	&\times& \left[\frac{2-N_c/C_F}{\alpha(y_1+\bar{y}_2)
 		+y_1\bar{y}_2}
 	-\frac{N_c/C_F}{y_2 \alpha+y_1(y_2-y_1)}\right]M(y_1,y_2;\mu^2)
 	\nonumber \\
  	 &+& \int^{1}_{0}dy_2\int^{y_2}_{0}dy_1 \Big[ {2+N_c/C_F \over \bar{y}_1}
 	+ {y_1 \over \alpha+y_1\bar{y}_1} 
 	\left({(2-N_c/C_F)y_1\alpha \over
 	\alpha(y_1+\bar{y}_2) + y_1\bar{y}_2}-2\right) \nonumber \\
  	&-&\frac{N_c}{C_F}\frac{(y_2-y_1)\bar{y}_2}{\bar{y}_1}\frac{1}{\alpha\bar{y}_1+(y_2-y_1)\bar{y}_2}\Big]\, S(y_1,y_2;\mu^2) \Big\}\,,
 	\label{Phi_TT}
 \end{eqnarray}
 where:
 \begin{equation}
C_F=\frac{N_c^2-1}{2N_c}\,,
 \end{equation}
 \begin{equation}
B(y_1,y_2;\mu^2)  =-5040 y_1 \bar{y}_2 (y_1-\bar{y}_2) (y_2-y_1)\,,
 \end{equation}
\begin{equation}
D(y_1,y_2;\mu^2)  =-360 y_1\bar{y}_2(y_2-y_1)
\left(1+\frac{\omega^{A}_{\{1,0\}}(\mu^2)}{2}\left(7\left(y_2-y_1\right)-3\right)
\right)\,,
\end{equation} 
and where the three-body DAs read:
\begin{equation}
 	M(y_1,y_2;\mu^2) = \zeta^{V}_{3V}(\mu^2) B(y_1,y_2;\mu^2) - \zeta^{A}_{3V}(\mu^2) D(y_1,y_2;\mu^2)\,,
 \end{equation}
 \begin{equation}
 	S(y_1,y_2;\mu^2) = \zeta^{V}_{3V}(\mu^2) B(y_1,y_2;\mu^2) + \zeta^{A}_{3V}(\mu^2) D(y_1,y_2;\mu^2)\,
\end{equation}
with the dimensionless coupling constants $\zeta^{V}_{3V}(\mu^2)$ and $\zeta^{A}_{3V}(\mu^2)$ defined as
\begin{equation}
\label{zeta}
	\zeta^{V}_{3V}(\mu^2) = \frac{f^{V}_{3V}(\mu^2)}{f_V}\,, \qquad 	\zeta^{A}_{3V}(\mu^2) = \frac{f^{A}_{3V}(\mu^2)}{f_V}\,.
\end{equation}
The dependence on the factorization scale $\mu^2$ can be determined from evolution equations~\cite{Ball:1998sk} (see also Appendix B in Ref.~\cite{Anikin:2011sa}), with the initial condition at a renormalization scale $\mu_0 = 1$ GeV.\\
The DAs $\varphi^T_1(y;\mu^2)$ and $\varphi^T_A(y;\mu^2)$ in Eq.~\eqref{Phi_TT}
encompass both genuine twist-3 and Wandzura-Wilczek~(WW)
contributions\footnote{Genuine terms are related to
	$B(y_1,y_2;\mu^2)$ and $D(y_1,y_2;\mu^2)$; WW contributions, instead, are those obtained
	in the approximation in which $B(y_1,y_2;\mu^2)=D(y_1,y_2;\mu^2)=0$. For their expressions in this last case see Eq.~(9) in Refs.~\cite{Anikin:2011sa, Bolognino:2018}.}~\cite{Ball:1998sk}. 

\subsection{Light-cone wave function (LCWF) approach}

In the light-cone $k_T$-factorization approach, the calculation proceeds in a 
slightly different way. Here one calculates the amplitude for the 
$ \gamma^*\,p \to q \bar q p$ diffractive process and projects the 
final state $q \bar q$ pair onto the vector meson state.
We treat the $\phi$-meson as a pure $s \bar s$ state. The meson of momentum $P= (P_+, m_\phi^2 /(2 P_+), {\bf{0}})$ is described by the $s \bar s$ light cone wave function as 
\begin{eqnarray}
\ket{\phi, P_+, \lambda_V} = \int {dy d^2 \bk \over y \bar y} \, \Psi^{(\lambda_V)}_{\lambda \bar \lambda}(y,\bk) \, \ket{ s(yP_+,\bk,\lambda)  \bar s(\bar yP_+,-\bk,\bar \lambda)} + \dots
\end{eqnarray} 

The amplitude for diffractive vector meson production then takes the form
\begin{eqnarray}
\Im m   T_{\lambda_V,\lambda_\gamma} (s,Q^2) 
= s \, \int {dy d^2 \bk \over y \bar y 16 \pi^3} \sum_{\lambda \bar \lambda} {\cal M}^{(\lambda_\gamma)}_{\lambda \bar \lambda}(\gamma^*p \to s \bar s p) \Psi^{(\lambda_V)*}_{\lambda \bar \lambda} (y,\bk) \, .
\end{eqnarray} 
The explicit expressions for the diffractive amplitudes can be found in Ref.~\cite{Ivanov:2004ax}. Here we are interested only in the forward scattering limit of vanishing transverse momentum transfer, where only the helicity conserving amplitudes with $\lambda_V = \lambda_\gamma$ contribute.

We can easily read off the following expressions for the impact factors of interest.
The $L \to L$ IF reads
\begin{eqnarray}
\Phi^{\gamma^* \to \phi}_{0,0}(\bkappa^2,Q^2) 
&=& \sqrt{4 \pi \alpha_{\rm em}} e_q \, 8 \pi  \alpha_S(\mu^2) Q \int {dy d^2 \bk \over \sqrt{y \bar y} 16 \pi^3} I_0(\bk,\bkappa) 
\nonumber \\
&\times& 
y \bar y \Big\{ \Psi^{(0)*}_{+-}(y,\bk) +  \Psi^{(0)*}_{-+}(y,\bk) \Big \}\,.
\end{eqnarray}
For the $T \to T$ IF we obtain 
\begin{eqnarray}
\Phi^{\gamma^* \to \phi}_{\pm,\pm} (\bkappa^2,Q^2) &=&  
\sqrt{4 \pi \alpha_{\rm em}} e_q
 4 \pi \alpha_S(\mu^2)  \int {dy d^2 \bk \over \sqrt{y \bar y} 16 \pi^3}  \nonumber \\
&\times& 
\Big[ (\be(\pm) \cdot \bI_1 (\bk,\bkappa))
\Big\{ (y - \bar y) \Big( \Psi^{(\pm)*}_{+-}(y,\bk) +  \Psi^{(\pm)*}_{-+}(y,\bk) \Big) 
\nonumber \\
&& + \Psi^{(\pm)*}_{+-}(y,\bk) -  \Psi^{(\pm)*}_{-+}(y,\bk)\Big\} 
+ \sqrt{2} m_q I_0(\bk,\bkappa)  \Psi^{(\pm)*}_{++}(y,\bk) \Big]
\, .
\nonumber \\
\end{eqnarray}
Here 
\begin{eqnarray}
I_0(\bk,\bkappa) = {1 \over \bk^2 + \varepsilon^2} - {1 \over (\bk+ \bkappa)^2 + \varepsilon^2} , \quad
\bI_1 (\bk,\bkappa) = {\bk \over \bk^2 + \varepsilon^2} - {\bk + \bkappa \over (\bk+ \bkappa)^2 + \varepsilon^2}, 
\end{eqnarray}
and $\varepsilon^2 = m_q^2 + y \bar y Q^2$.
We now want to compare these results with the twist expansion
approach presented in the previous chapters.
To this end, we should expand the impact factors around the 
limit of collinear kinematics for the $q \bar q$-pair.
While an analogous expansion around the small-$\bkappa$ limit,
has been discussed in great detail, the analogous comparison to
leading and higher twist distribution amplitudes is up to now 
missing. 

Expanding in $\bk^2/(\bkappa^2 + \varepsilon^2) \ll 1$, we obtain
\begin{eqnarray}
I_0(\bk,\bkappa) \approx {1 \over \varepsilon^2 } 
- {1 \over \bkappa^2 + \varepsilon^2} = 
{\bkappa^2 \over \varepsilon^2 (\bkappa^2 + \varepsilon^2)} \, ,
\end{eqnarray}
and 
\begin{eqnarray}
\bI_1 (\bk,\bkappa) \approx \bk {\bkappa^2 \over \varepsilon^2 (\bkappa^2 + \varepsilon^2)}  + {2 (\bk \cdot \bkappa) \bkappa \over (\bkappa^2 + \varepsilon^2)^2} \to {\bkappa^2 (\bkappa^2 + 2 \varepsilon^2) \over \varepsilon^2 (\bkappa^2 + \varepsilon^2)^2} \bk \, ,
\end{eqnarray}
where we performed the azimuthal average in the last step.
Inserting the expanded $I_0$ into the LL IF, we find
\begin{eqnarray}
\Phi^{\gamma^* \to \phi}_{0,0}(\bkappa^2,Q^2) 
&=& \sqrt{4 \pi \alpha_{\rm em}} e_q \, 8 \pi  \alpha_S(\mu^2) Q
\int_0^1 dy \, y \bar y {\bkappa^2 \over \varepsilon^2 (\bkappa^2 + \varepsilon^2)} \,\nonumber \\
&\times&  {1 \over \sqrt{y \bar y}} \int{d^2 \bk \over 16 \pi^3}
\Big\{ \Psi^{(0)*}_{+-}(y,\bk) +  \Psi^{(0)*}_{-+}(y,\bk) \Big \} \theta(\mu^2 - \bk^2) \, \nonumber \\
&=& \sqrt{4 \pi \alpha_{\rm em}} e_q { 4 \pi \alpha_S(\mu^2) f_V  \over N_c Q } \int_0^1 dy { y \bar y \over (y \bar y + \tau)} {\alpha \over (\alpha + y \bar y + \tau)} \, \varphi_1(y,\mu^2) \, . \nonumber \\
\label{IF_LL_expI0}
\end{eqnarray}
Here we introduced the variables $\alpha = \bkappa^2/Q^2$ and
$\tau = m_q^2/Q^2$. We see that we have obtained a generalization
to finite quark mass of the impact factor of Eq.~\eqref{Phi_LL}.
The helicity combination of the LCWF which appears under the $\bk$ integral gives rise to the leading twist distribution amplitude 
of the longitudinally polarized vector meson, defined following the rules of Ref.~\cite{Lepage:1980fj} as
\begin{eqnarray}
f_V \varphi_1(y,\mu_0^2) = {2 N_c \over \sqrt{y \bar y}} \, \int {d^2\bk \over 16 \pi^3} \theta(\mu_0^2 - \bk^2)
\Big\{ \Psi^{(0)*}_{+-}(y,\bk) +  \Psi^{(0)*}_{-+}(y,\bk) \Big \}\,. 
\end{eqnarray} 
The scale $\mu^2$ in Eq.~\eqref{IF_LL_expI0} must be chosen such that the small-$\bk$ expansion is valid, i.e. $\mu^2 \sim (Q^2 + m_\phi^2)/4$.

We can now follow a similar strategy for the transverse IF.
To that end we introduce the following representations of the
higher twist DA's: 
\begin{eqnarray}
f_V \varphi_1^T(y,\mu_0^2) &=& {2 N_c \over \sqrt{y \bar y}  }
\int {d^2\bk \over 16 \pi^3} \theta(\mu_0^2 - \bk^2)
(\be(\pm) \cdot \bk)\Big\{ \Psi^{(\pm)*}_{+-}(y,\bk) +  \Psi^{(\pm)*}_{-+}(y,\bk) \Big \}\,, \nonumber \\
f_V \varphi_A^T(y,\mu_0^2) &=& {2 N_c \over \sqrt{y \bar y}  }
\int {d^2\bk \over 16 \pi^3} \theta(\mu_0^2 - \bk^2)
(\be(\pm) \cdot \bk)\Big\{ \Psi^{(\pm)*}_{+-}(y,\bk) -  \Psi^{(\pm)*}_{-+}(y,\bk) \Big \}\,, \nonumber \\
f_V \varphi_m(y,\mu_0^2) &=& 	{2 N_c \over \sqrt{y \bar y}  }
\int {d^2\bk \over 16 \pi^3} \theta(\mu_0^2 - \bk^2) \sqrt{2} m_q \Psi^{(\pm)*}_{++}(z,\bk)\,.
\end{eqnarray}
We notice, that
\begin{eqnarray}
\int_0^1 dy \, \varphi(y,\mu_0^2) = 1, \qquad \int_0^1 dy \, \varphi_1^T(y,\mu_0^2) = 0. 
\end{eqnarray}
The transverse IF that we derive is again a massive generalization of Eq.~\eqref{Phi_TT} and reads
\begin{eqnarray}
\Phi^{\gamma \to \phi}_{\pm,\pm}(\bkappa^2,Q^2) &=& \sqrt{4 \pi \alpha_{\rm em}} e_q { 2 \pi \alpha_S(\mu^2) f_V  \over N_c Q^2 }
\int_0^1 {dy \over y \bar y + \tau} \Big\{ {\alpha (\alpha + 2 y \bar y + 2 \tau) \over (\alpha + y \bar y +\tau)^2} \nonumber \\
&\times& \hspace{-0.2cm}
\Big( (y - \bar y) \varphi_1^T(y,\mu^2) + \varphi_1^A(y,\mu^2) 
\Big) 
+ {\alpha \over \alpha + y \bar y + \tau} \varphi_m(y,\mu^2)
\Big \}.
\end{eqnarray}
We realize that up to the DA $\varphi_m$, which vanishes in the massless limit, the structure of the IF is exactly the same
as for the one of Eq.~\eqref{Phi_TT} neglecting the so-called genuine three particle distributions. The latter obviously
can appear only at the level of the $q \bar q g$-Fock state.

We now wish to give some explicit expressions for the DA's in
question. To this end, we use the $V \to q \bar q$ vertex
from Ref. \cite{Ivanov:2004ax}, where the $\phi$-meson is treated as a pure $S$-wave bound state of strange quark and antiquark. 
For the relevant combinations of light-cone wave functions we obtain in the case of the longitudinally polarized vector meson:
\begin{eqnarray}
\Psi^{(0)*}_{+-}(y,\bk) +  \Psi^{(0)*}_{-+}(y,\bk) = - 4M \sqrt{y \bar y} \Big\{ 1 + {(y- \bar y)^2 \over 4 y \bar y} {2 m_q \over M + 2 m_q}   
\Big\} \psi(y,\bk) \, .
\end{eqnarray}
The radial wave function $\psi(z,\bk)$ is normalized as
\begin{eqnarray}
N_c \int {dz d^2 \bk \over y \bar y 16 \pi^3} 2 M^2 \, |\psi(y,\bk)|^2 = 1\, .
\end{eqnarray}
Above $M^2 = (\bk^2 + m_q^2)/(y \bar y)$ is the invariant mass of the $s \bar s$-system.
We can now express the leading twist DA through the radial WF as
\begin{eqnarray}
f_V \varphi_1(y,\mu_0^2) = {N_c \over 2 \pi^2}  \int_0^{\mu_0^2}  d\bk^2 
M \Big\{ 1 + {(y - \bar y)^2 \over 4 y \bar y} {2 m_q \over M + 2 m_q}   
\Big\} \psi(y,\bk) \, .
\end{eqnarray}
Now, for the higher twist DA's of the transversely polarized vector meson, we obtain
\begin{eqnarray}
f_V \varphi_1^T(y,\mu_0^2) &=& (y - \bar y){N_c \over 8 \pi^2}  \int_0^{\mu_0^2}  d\bk^2 \, \bk^2 
 {M \over M + 2 m_q} {\psi(y,\bk) \over y \bar y} \,, \nonumber \\
f_V \varphi_A^T(y,\mu_0^2) &=& {N_c \over 4 \pi^2}  \int_0^{\mu_0^2}  d\bk^2 \, \bk^2  {\psi(y,\bk) \over y \bar y} \,, \nonumber \\ 
f_V \varphi_m(y,\mu_0^2) &=& m_q^2 {N_c \over 4 \pi^2}  \int_0^{\mu_0^2}  d\bk^2 \, 
\Big\{ 1 + {\bk^2 \over m_q (M+2 m_q)} \Big \} {\psi(y,\bk) \over y \bar y} \,. \nonumber \\ 
\end{eqnarray}

\subsection{Characteristic parameters}

Typical constants for $\phi$ meson, entering DAs and IFs, used in numerical computations, are provided in the following tables:
\begin{table}[htb]
	$$
	\renewcommand{\arraystretch}{1.4}
	\addtolength{\arraycolsep}{3pt}
	\begin{array}{|c|c|}
	\hline
	V & \phi\\ \hline
	f_V [{\rm GeV}] & 0.254\\
	\zeta_{3V}^A & 0.032\\
	\zeta_{3V}^V  & 0.013\\
	\omega_{1,0}^A & -2.1 \\
	\omega_{1,0}^V & 28/3 \\
	\hline
	\end{array}
	$$
	\caption[]{Experimental value of coupling to the vector
		current \protect{\cite{PDG}}\label{tab:B1} (first row); couplings entering the vector meson DAs at the scale $\mu_0 = 1$ GeV.}
	\renewcommand{\arraystretch}{1.4}
	\addtolength{\arraycolsep}{3pt}
 \end{table}
 \begin{table}[htb]
 	$$
 	\renewcommand{\arraystretch}{1.4}
 	\addtolength{\arraycolsep}{3pt}
 	\begin{array}{|c|c|}
 	\hline
 	V & \phi\\ \hline
 	m_Vf^{A}_{3V}[{\rm GeV^2}] & 3.37\cdot10^{-3} \\
 	m_Vf^{V}_{3V}[{\rm GeV^2}] & 5.26\cdot10^{-3} \\
 	\hline
 	\end{array}
 	$$
 	\caption[]{Decay constants obtained from Eq.~\eqref{zeta}.}
 	\renewcommand{\arraystretch}{1.4}
 	\addtolength{\arraycolsep}{3pt}
 \end{table}	\renewcommand{\arraystretch}{1.4}

 \subsection{Cross section and b-slope}

 The imaginary part of the amplitude in Eq.~\eqref{amplitude} which enters the expression of the cross section for transverse and longitudinal polarization, can be written as:
 \begin{equation}
 	\sigma_L\,(\gamma^*\,p \rightarrow V\,p) = \frac{1}{16 \pi B(Q^2)}\left|\frac{T_{00}(s,Q^2)}{W^2}\right|^2\,,
 \end{equation}
 \begin{equation}
 \sigma_T\,(\gamma^*\,p \rightarrow V\,p) = \frac{1}{16 \pi B(Q^2)}\left|\frac{T_{11}(s,Q^2)}{W^2}\right|^2\,,
 \end{equation}
 where $B(Q^2)$ is a slope parameter which depends on the virtuality of the photon and it is parametrized in the present analysis as follows~\cite{Nemchik:1997xb}:
 \begin{equation}
 \label{slope_B}
 	B(Q^2) = \beta_0 - \beta_1\,\log\left[\frac{Q^2+m_\phi^2}{m^2_{J/\psi}}\right]+\frac{\beta_2}{Q^2+m_\phi^2}\,,
 \end{equation}
 with $\beta_0 = 7.0$ GeV$^{-2}$, $\beta_1 = 1.1$  GeV$^{-2}$ and $\beta_2 = 1.1$.\\
\begin{figure}[htb]
	\centering	
	\includegraphics[scale=0.30,clip]{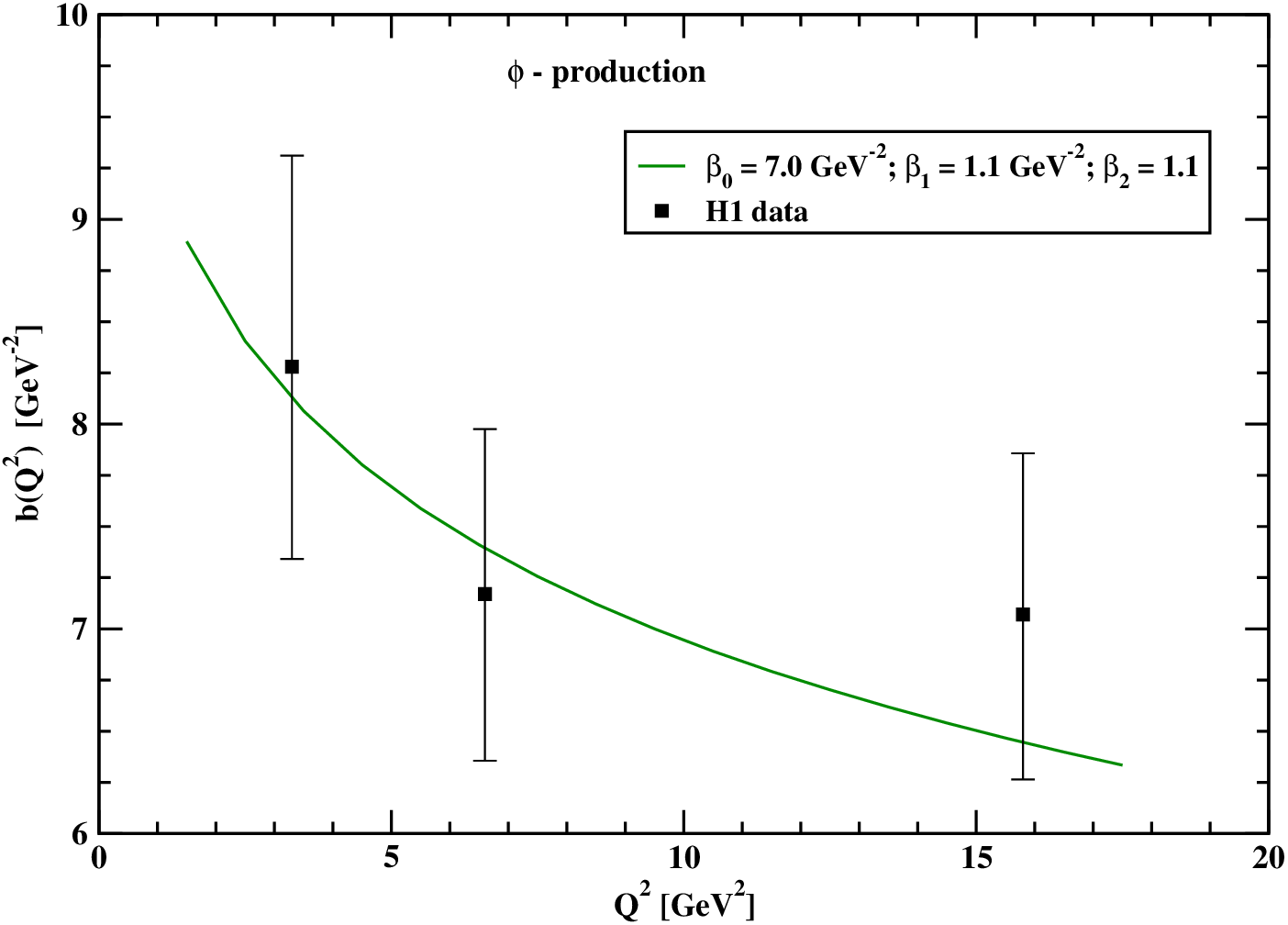}
	\caption{$Q^2$-dependence of the b-slope for $\phi$-meson production in the $\gamma^*\,p \to \phi\,p$ reaction. Due to the high uncertainty of the experimental data, we keep the standard choice of $\beta_0$, $\beta_1$ and $\beta_2$ parameters from Ref.~\cite{Nemchik:1997xb} for all our results.}
	\label{fig:bslope}
\end{figure} 
 The full cross section is a sum of longitudinal and transverse
 components, and it reads
 \begin{equation}
 	\sigma_{tot} (\gamma^*\,p \rightarrow V\,p) = \sigma_T + \epsilon\,\sigma_L\,,
 \end{equation}
 where $\epsilon \approx 1$ due to HERA kinematics.

 \section{Numerical Results}
\begin{figure}[htb]
	\centering
	\includegraphics[scale=0.30,clip]{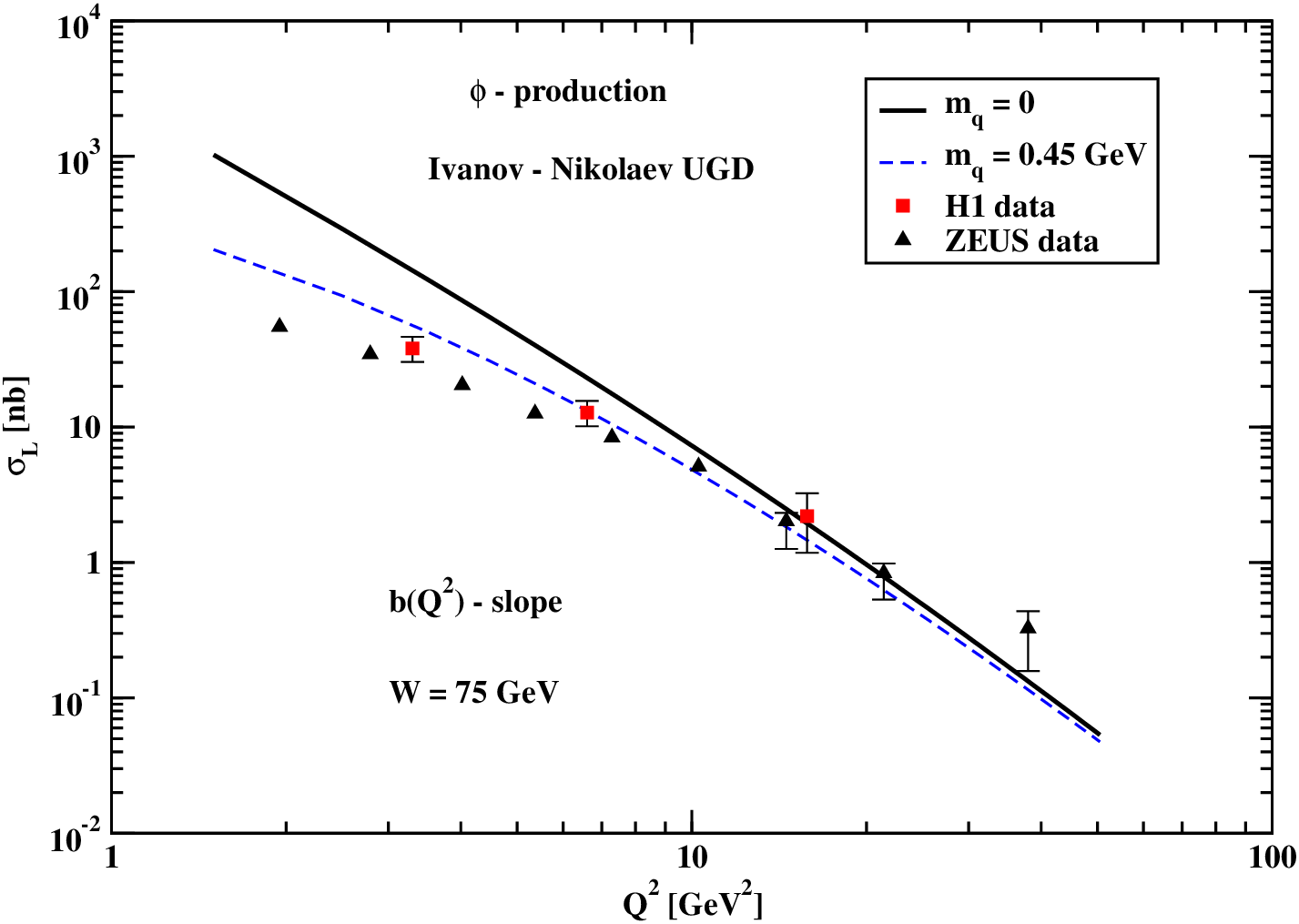}
	\includegraphics[scale=0.30,clip]{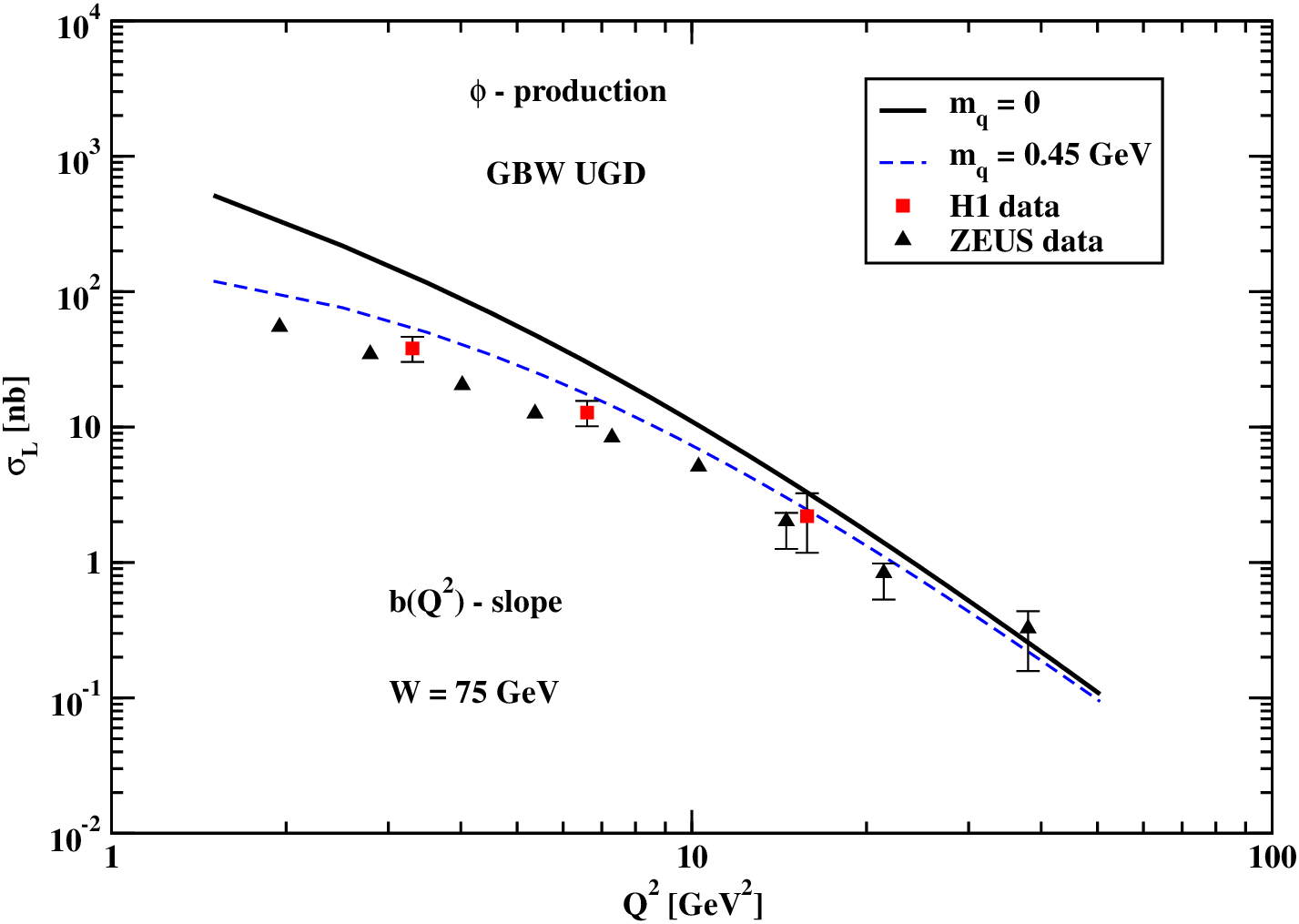}
	\caption{$Q^2$-dependence of longitudinal cross section $\sigma_L$ of $\phi$-meson production, at $W = 75$~GeV, in comparison with experimental data of the H1~\cite{Aaron:2009xp} and ZEUS \cite{ZEUScoll2005} collaborations. The result is obtained within the $k_T$-factorization using the Ivanov-Nikolaev UGD model (left panel) and the GBW one (right panel). The solid lines are for the case when the strange-quark mass is neglected and the dashed lines for the strange-quark mass fixed at $m_q = 0.45$ GeV. In both cases, the asymptotic distribution amplitude (DA) is used.}
	\label{fig:sigmaL_IN_DAs_mq}
\end{figure}

In what follows we present theoretical predictions adopting two different UGD models:
\begin{itemize}
\item the Ivanov-Nikolaev parametrization, endowed with soft and hard components to probe both large and small transverse momentum region (see Ref.~\cite{Ivanov:2000cm} for further details);
\item the model provided by Golec-Biernat and W\"usthoff (GBW), which derives from the dipole cross section for the scattering of a $q\bar{q}$ pair off a nucleon \cite{GolecBiernat:1998js}.
\end{itemize}
We start from calculating longitudinal cross section using 
the formalism described in the previous section.
In Fig.\ref{fig:sigmaL_IN_DAs_mq} we show results of our calculation
for the Ivanov-Nikolaev (left panel) and GBW (right panel) UGDs.
In order to get these predictions, the asymptotic DA has been used. This calculation has been obtained for $W$ = 75~GeV.
We observe that the cross sections obtained for massless strange quarks
(black solid line) overestimate the experimental cross section below $Q^2 <$ 10 GeV$^2$.
This is very different for $\rho^0$ production to be discussed elsewhere.
\begin{figure}[htb]
	\centering
	\includegraphics[scale=0.30,clip]{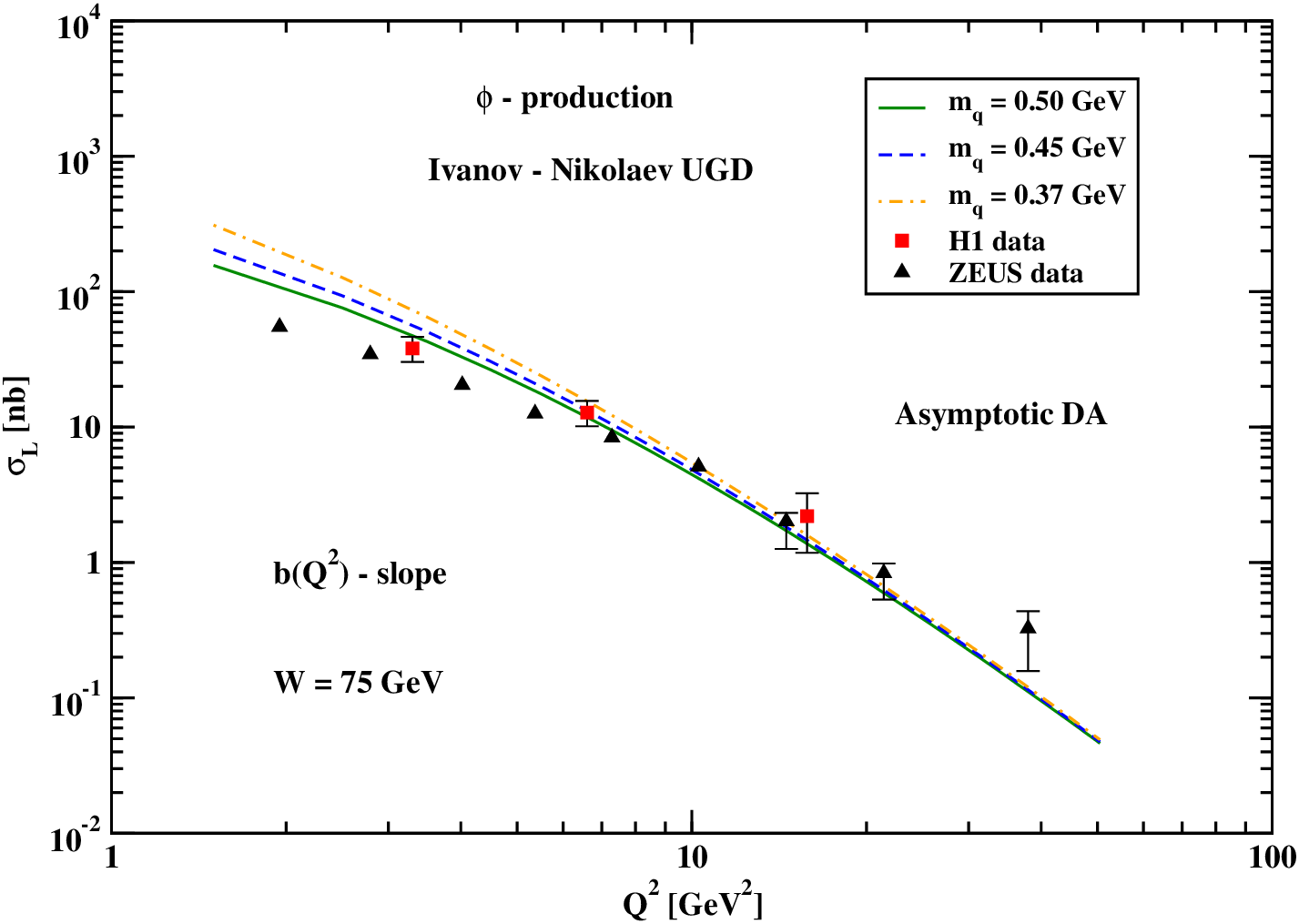}
	\includegraphics[scale=0.30,clip]{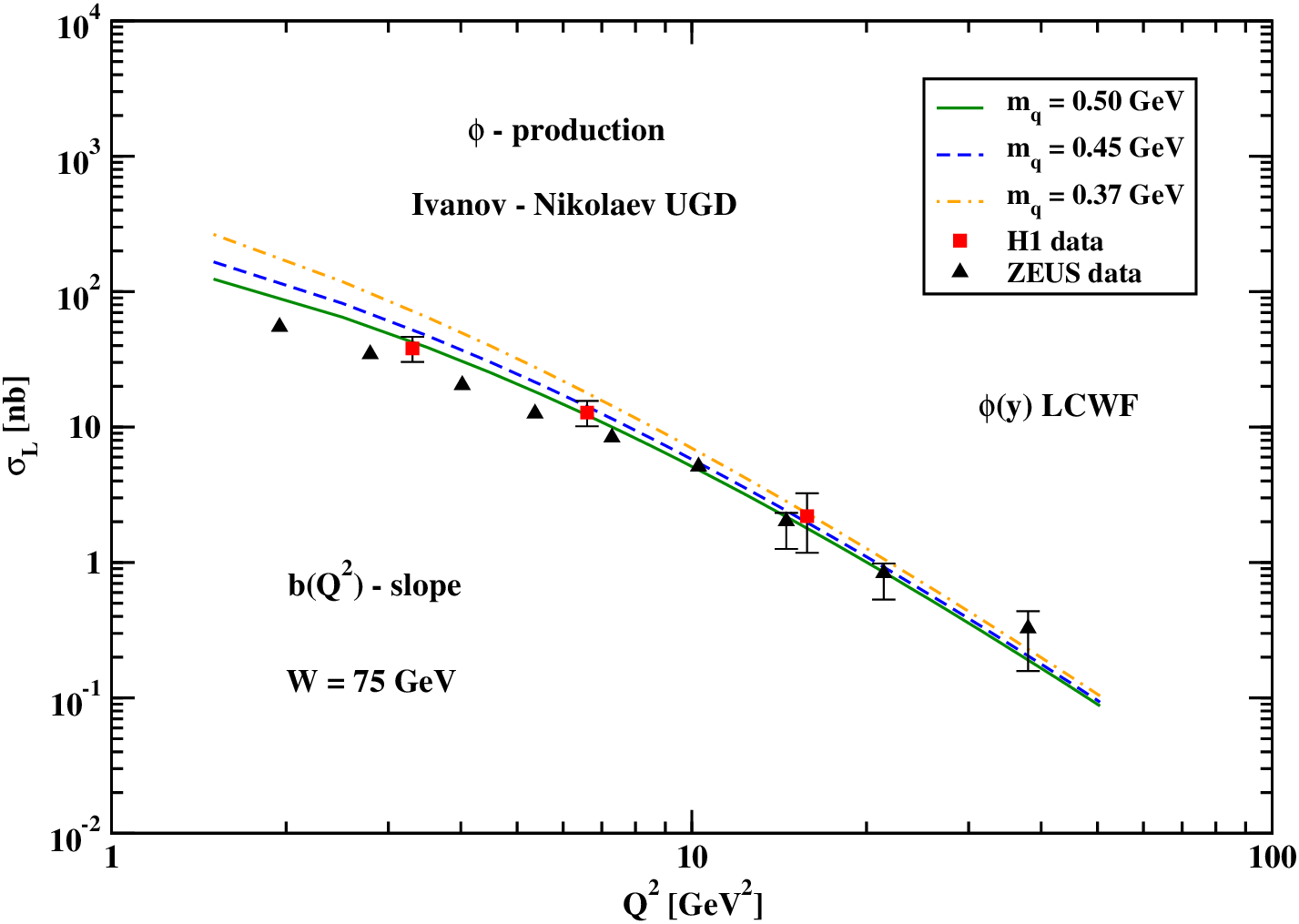}
	\caption{$\sigma_L$ for the asymptotic DA (left panel) and for the LCWF DA (right panel). Results for three different strange-quark-mass values are shown. Predictions are given using the Ivanov-Nikolaev UGD model.}
	\label{fig:systematics_sigL_asy_da}
\end{figure}
In both cases we present also our results when using quarks/antiquarks
with effective masses (as described in the previous section).
Then a good description of the experimental data is obtained for both the Ivanov-Nikolaev and GBW UGDs.
\begin{figure}[htb]
	\centering
	\includegraphics[scale=0.30,clip]{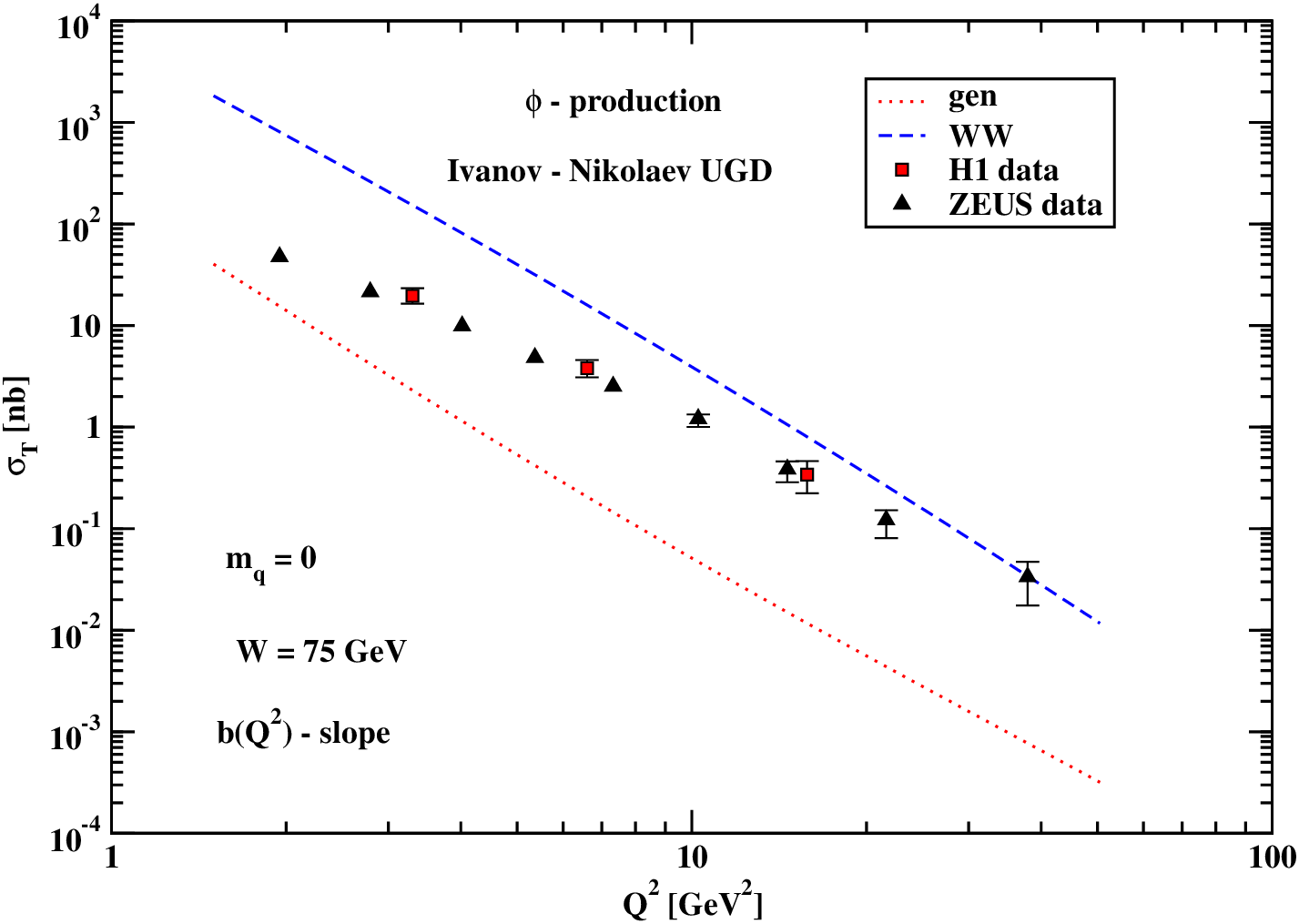}
	\caption{$Q^2$-dependence of the transverse cross section $\sigma_T$ neglecting the quark mass $m_q$. The WW and the genuine three-parton contributions are shown separately.}
	\label{fig:WW_GEN_sigT}
\end{figure}
How much the cross section depends on the quark mass is discussed in 
Fig.\ref{fig:systematics_sigL_asy_da}
for asymptotic (left panel) and LCWF (right panel) DAs, respectively.
The best description of the data is obtained with $m_q$ = 0.5 GeV.
A similar result was found in Ref.~\cite{Cisek:2010jk} within the $k_T$-factorization
approach with a Gaussian $s\bar{s}$ light-cone wave function for the $\phi$ meson.\\
Now we pass to the transverse cross section as a function of photon virtuality.

\begin{figure}[htb]
	\centering	
	\includegraphics[scale=0.30,clip]{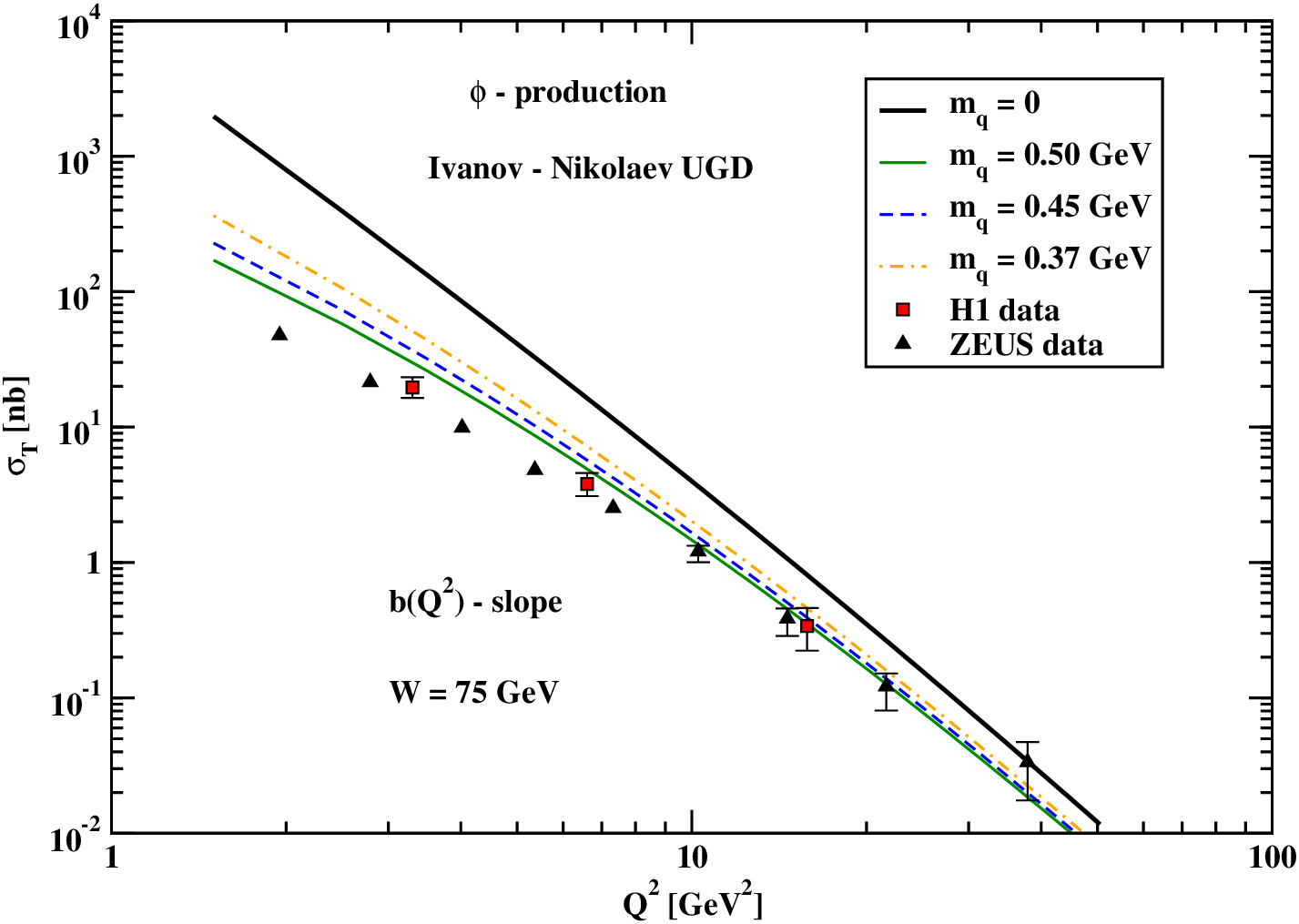}
	\includegraphics[scale=0.30,clip]{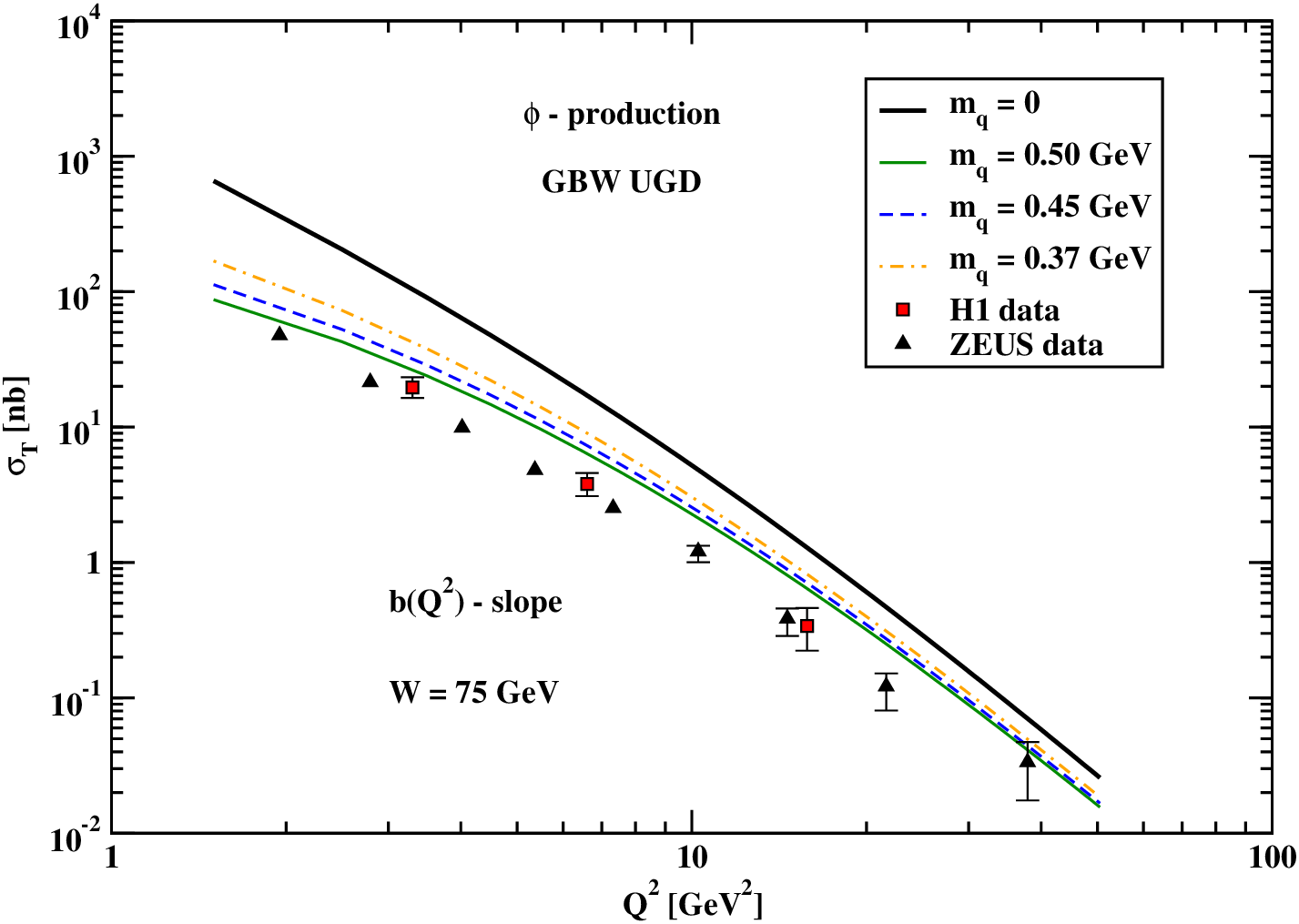}
	\caption{$Q^2$-dependence of the transverse cross section $\sigma_T$ for the $\phi$-meson production in the $\gamma^*\,p \to \phi\,p$ reaction, at $W = 75$~GeV, in comparison with the H1~\cite{Aaron:2009xp} and ZEUS~\cite{ZEUScoll2005} experimental data. The result is obtained within the $k_T$-factorization using the Ivanov-Nikolaev UGD model (left panel) and the GBW one (right panel). The thick solid line is for the case when the strange-quark mass is neglected. The thin lines show the result for the three different values of the strange-quark mass. In both cases, the curves were obtained with the asymptotic choice of the distribution amplitude (DA).}
	\label{fig:sigT_GBW_IN}
\end{figure}
\begin{figure}[htb]
  	\centering	
  	\includegraphics[scale=0.30,clip]{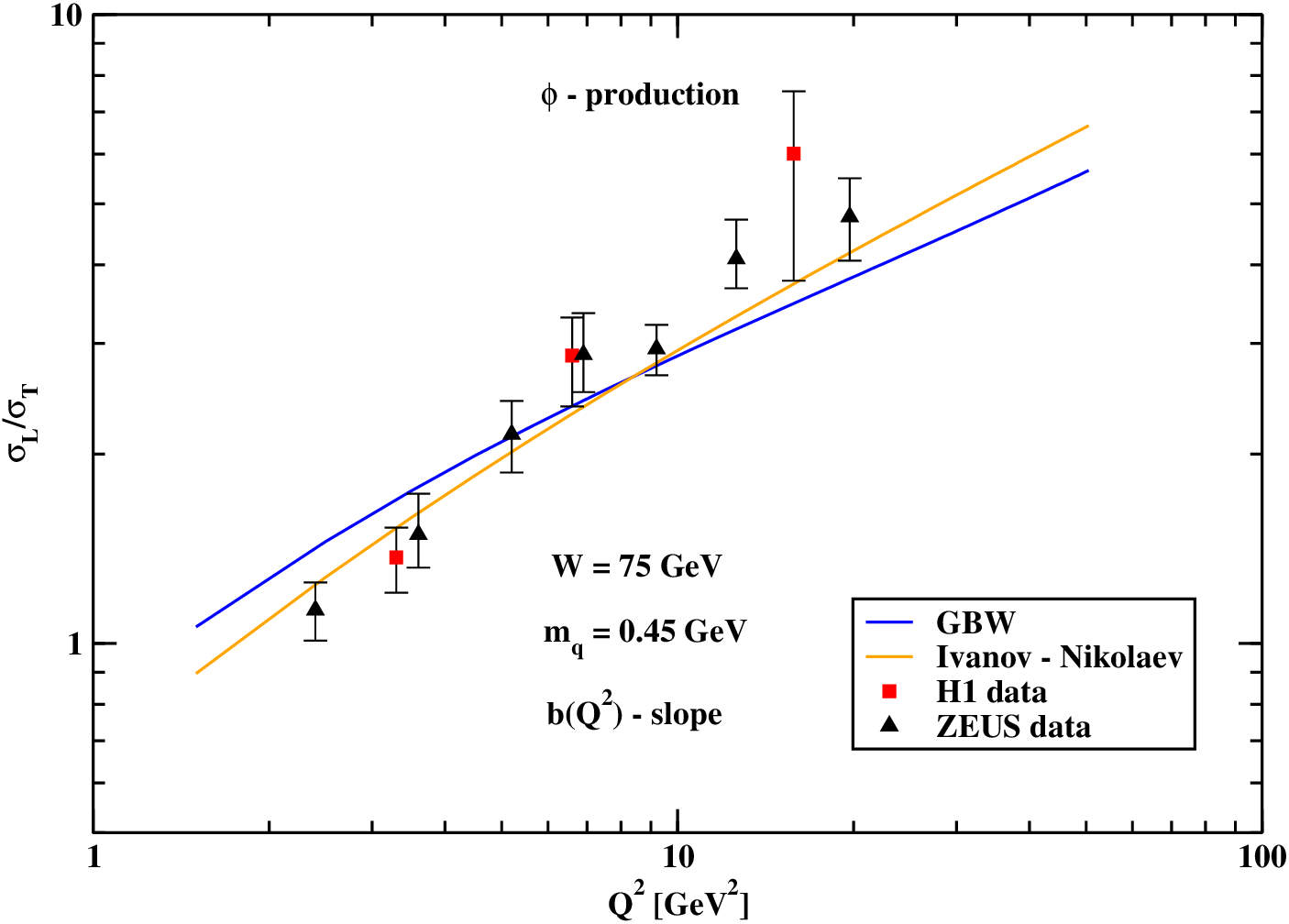}
  	\caption{$Q^2$-dependence of cross section ratio $\sigma_L/\sigma_T$ for the $\phi$-meson production in the $\gamma^*\,p \to \phi\,p$ reaction, at $W = 75$~GeV, in comparison with experimental data of the H1~\cite{Aaron:2009xp} and ZEUS~\cite{ZEUScoll2005} collaboration. The prediction is performed in the $k_T$-factorization approach, using both UGD models: the Ivanov-Nikolaev and the GBW one. The strange-quark mass is fixed here at $m_q = 0.45$ GeV.}
  	\label{fig:ratio_comparison_UGD}
  \end{figure}

In Fig.\ref{fig:WW_GEN_sigT} we show the cross section separately for the 
Wandzura-Wilczek and genuine three-parton contributions. In this calculation massless quarks
were used. We observe that the transverse cross section for the genuine
three-parton contribution is rather small. However, the WW contribution for massless quarks, 
similarly as for the longitudinal cross section, overpredicts the H1 and ZEUS
data. Can this be explained as due to the mass effect discussed in the
previous section?\\
In Fig.\ref{fig:sigT_GBW_IN} we show how the WW contribution changes when
including the mass effect discussed in the previous section.
Inclusion of the mass effect improves the description of the H1 and ZEUS experimental
transverse cross section. The description is, however, not perfect. In Fig.\ref{fig:sigT_GBW_IN} we present a similar result for both UGD models.
Unlike for the the longitudinal cross section, here the GBW overpredicts
the experimental data in the whole range of virtuality, while the Ivanov-Nikolaev model only at smaller $Q^2$ values. A reasonable result is obtained when including mass effect.\\
We wish to show also results for the $\sigma_L/\sigma_T$ ratio (see Fig.\ref{fig:ratio_comparison_UGD}) as a function of photon
virtuality $Q^2$ for the Ivanov-Nikolaev and GBW UGDs.
In this calculation the quark mass was fixed for $m_q$ = 0.45 GeV.
The Ivanov-Nikolaev UGD much better describes the H1 and ZEUS data.


\begin{figure}[htb]
	\centering
	\includegraphics[scale=0.30,clip]{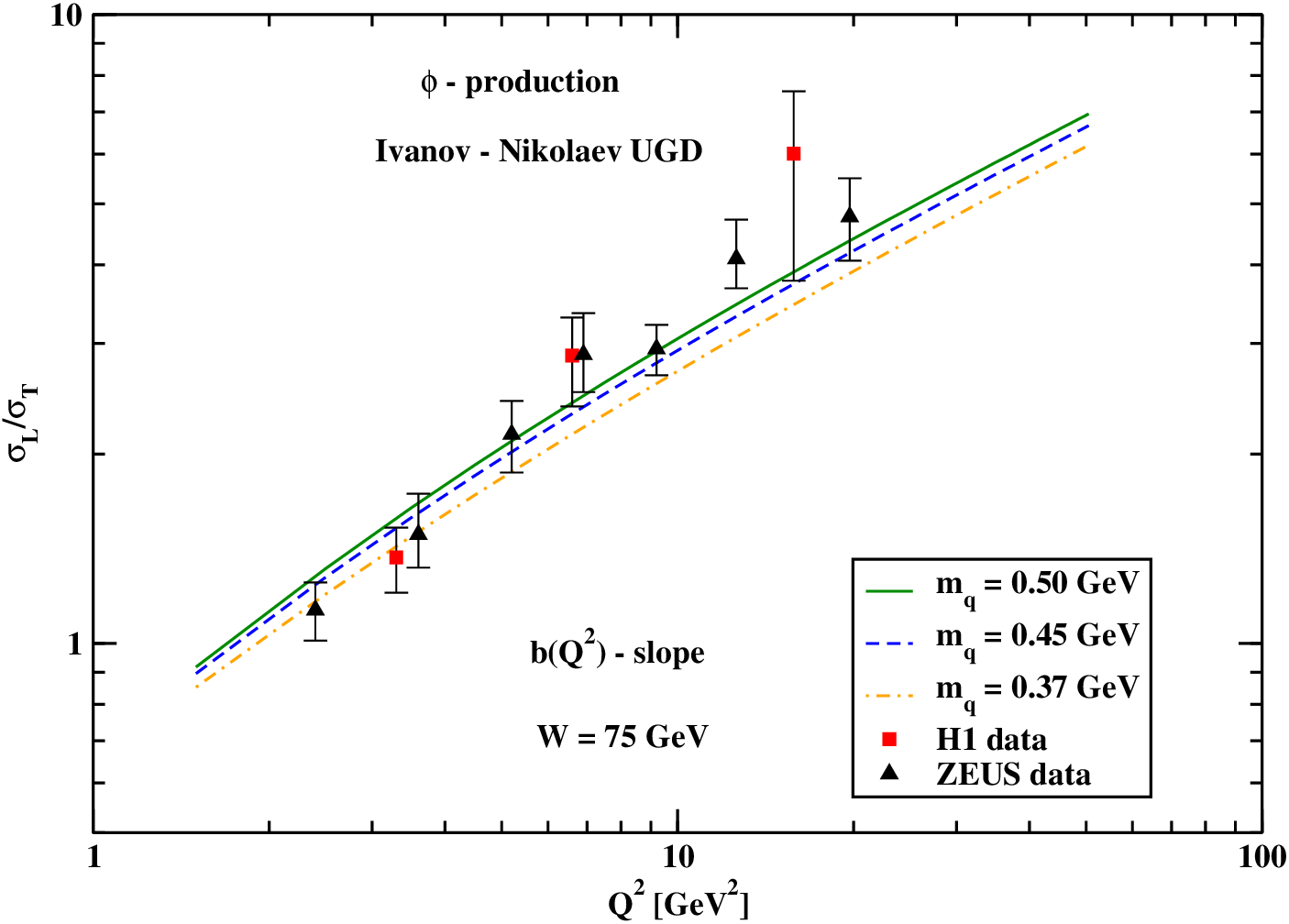}
	\includegraphics[scale=0.30,clip]{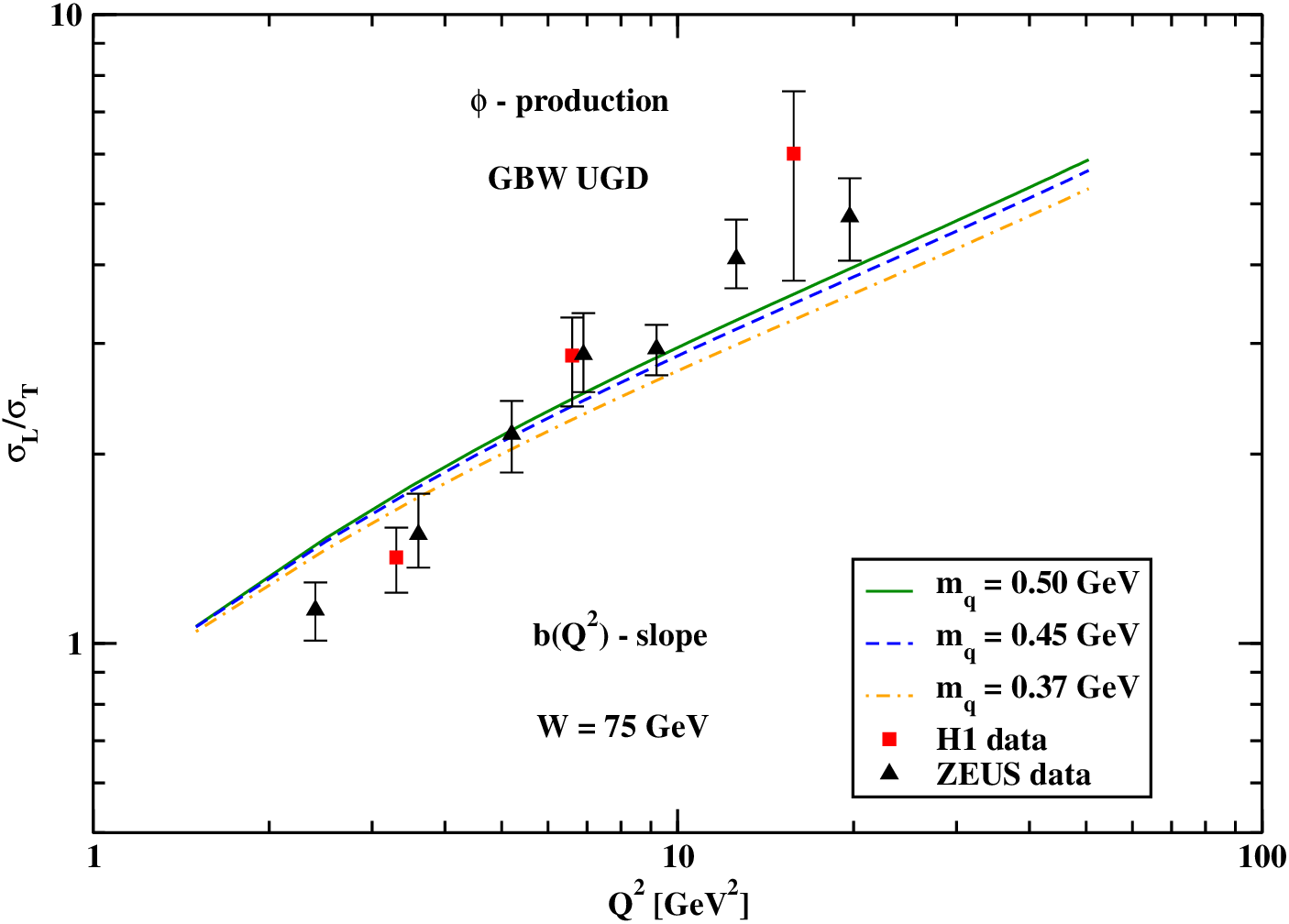}
	\caption{$Q^2$-dependence of the cross section ratio $\sigma_L/\sigma_T$ for three different quark-mass values $m_q$ using the Ivanov-Nikolaev (left panel) and the GBW (right panel) UGDs.}
	\label{fig:ratio_IN_comparison_masses}
\end{figure} 
How much the ratio depends on the effective quark mass parameter?
This is shown in Fig.\ref{fig:ratio_IN_comparison_masses}.
The ratio is much less sensitive to the quark mass than the polarized cross sections
$\sigma_L$ and/or $\sigma_T$ separately. So the extraction of the mass
parameter from the normalized cross section is preferred.

\begin{figure}[htb]
	\centering
	\includegraphics[scale=0.30,clip]{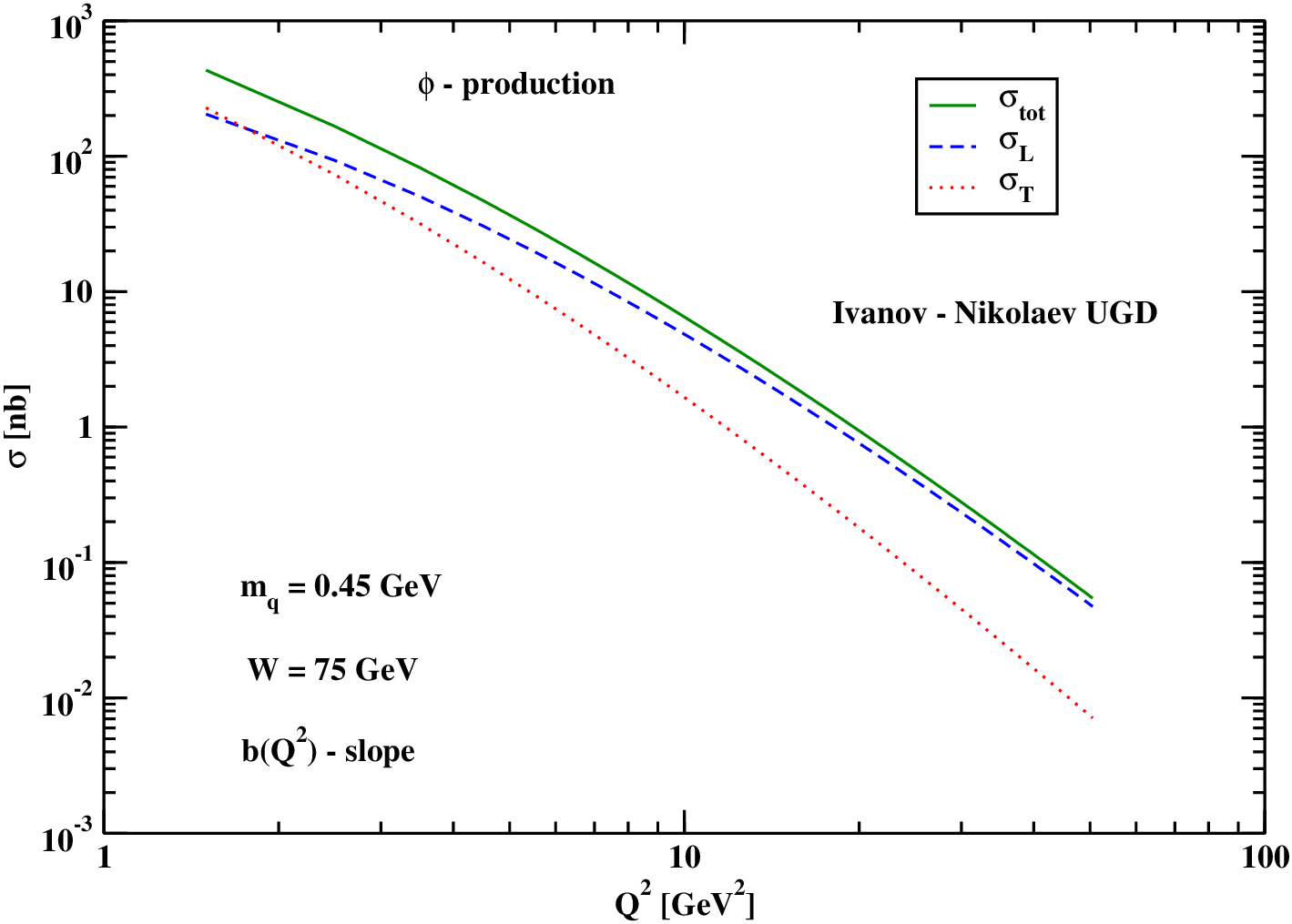}
	\includegraphics[scale=0.30,clip]{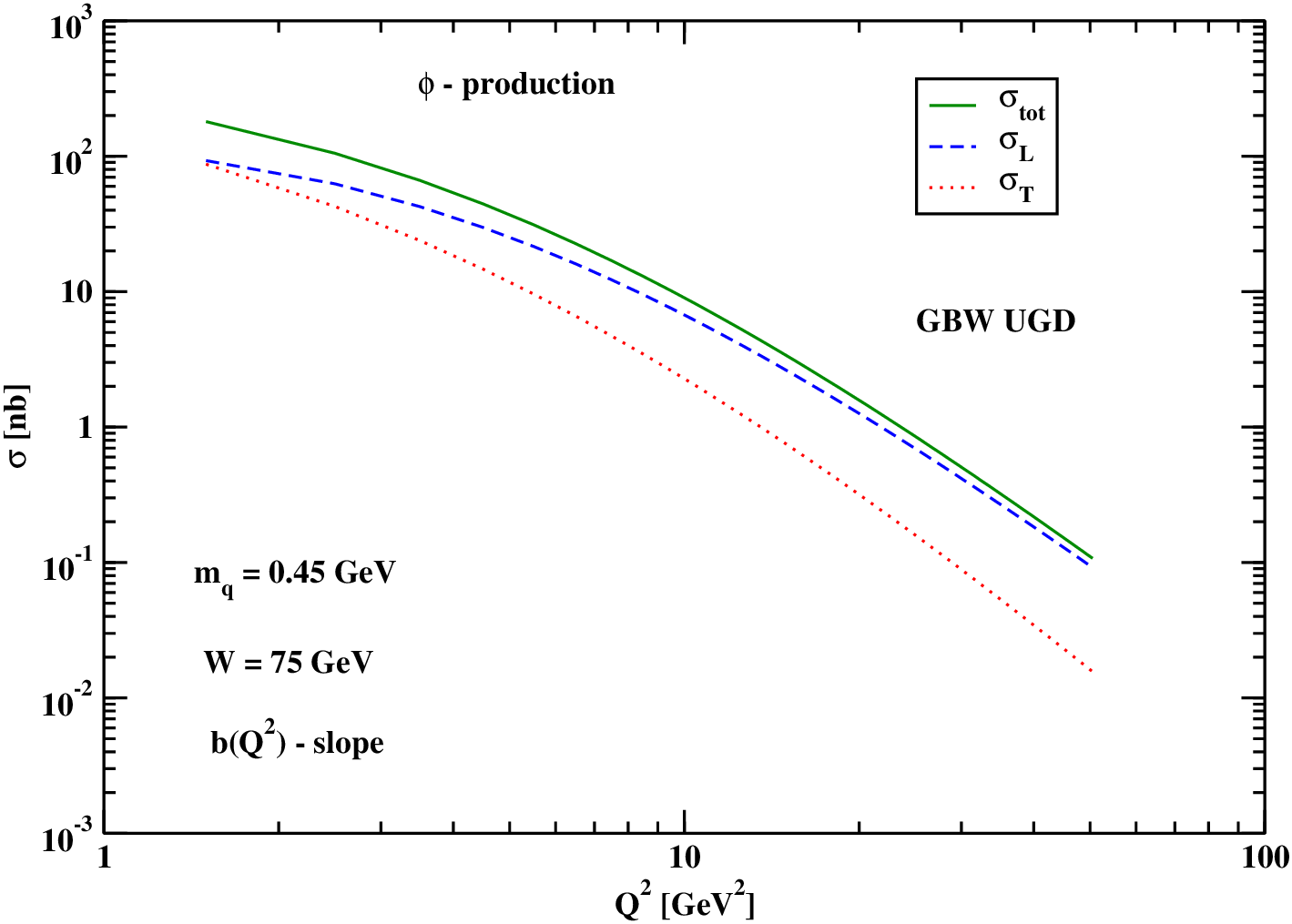}
	\caption{Longitudinal, transverse and total cross sections as functions of $Q^2$ using the Ivanov-Nikolaev (left panel) and the GBW (right panel) UGDs.}
	\label{fig:all_sigma_GBW_IN}
\end{figure}

\begin{figure}[htb]
	\centering
	\includegraphics[scale=0.30,clip]{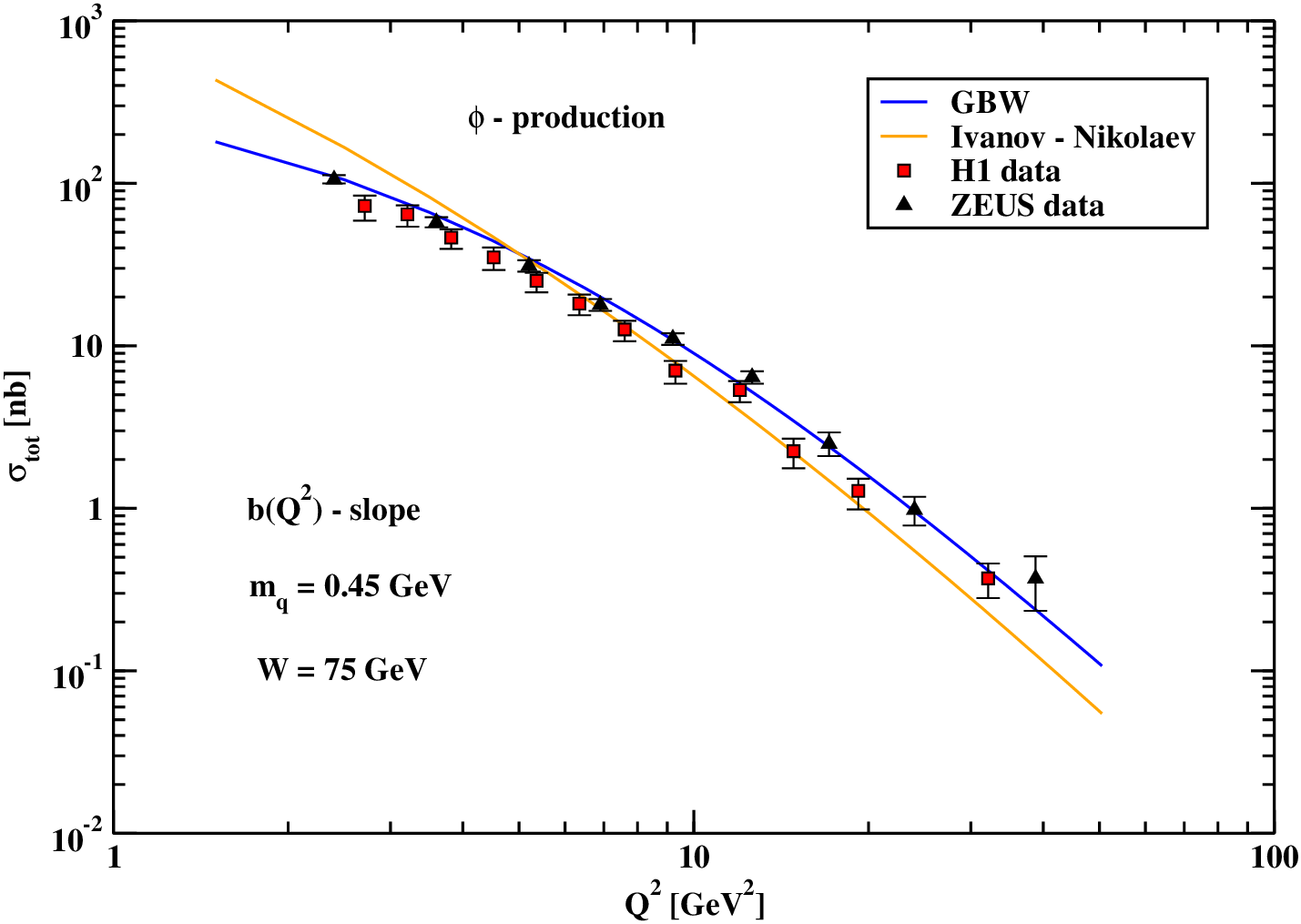}
	\caption{$Q^2$dependence of total cross section $\sigma_{\rm tot}$ at $W = 75$ GeV for both UGD models in comparison with the H1~\cite{Aaron:2009xp} and ZEUS~\cite{ZEUScoll2005} experimental data.}
	\label{fig:ZEUS_comparison}
\end{figure}

Now we shall show the total cross section $\sigma_{\rm tot}$ as a function of virtuality.
In Fig.~\ref{fig:all_sigma_GBW_IN} we show both longitudinal and transverse components as well as
their sum. The transverse cross section is somewhat steeper (falls
faster with virtuality) than the longitudinal one.
The comparison with the HERA data is presented in Fig.\ref{fig:ZEUS_comparison}.
The GBW UGD better describes the experimental data at small photon virtualities. There seems to be a small inconsistency of the
H1 and ZEUS data at larger virtualities.
  
\begin{figure}[htb]
\centering
\includegraphics[scale=0.30,clip]{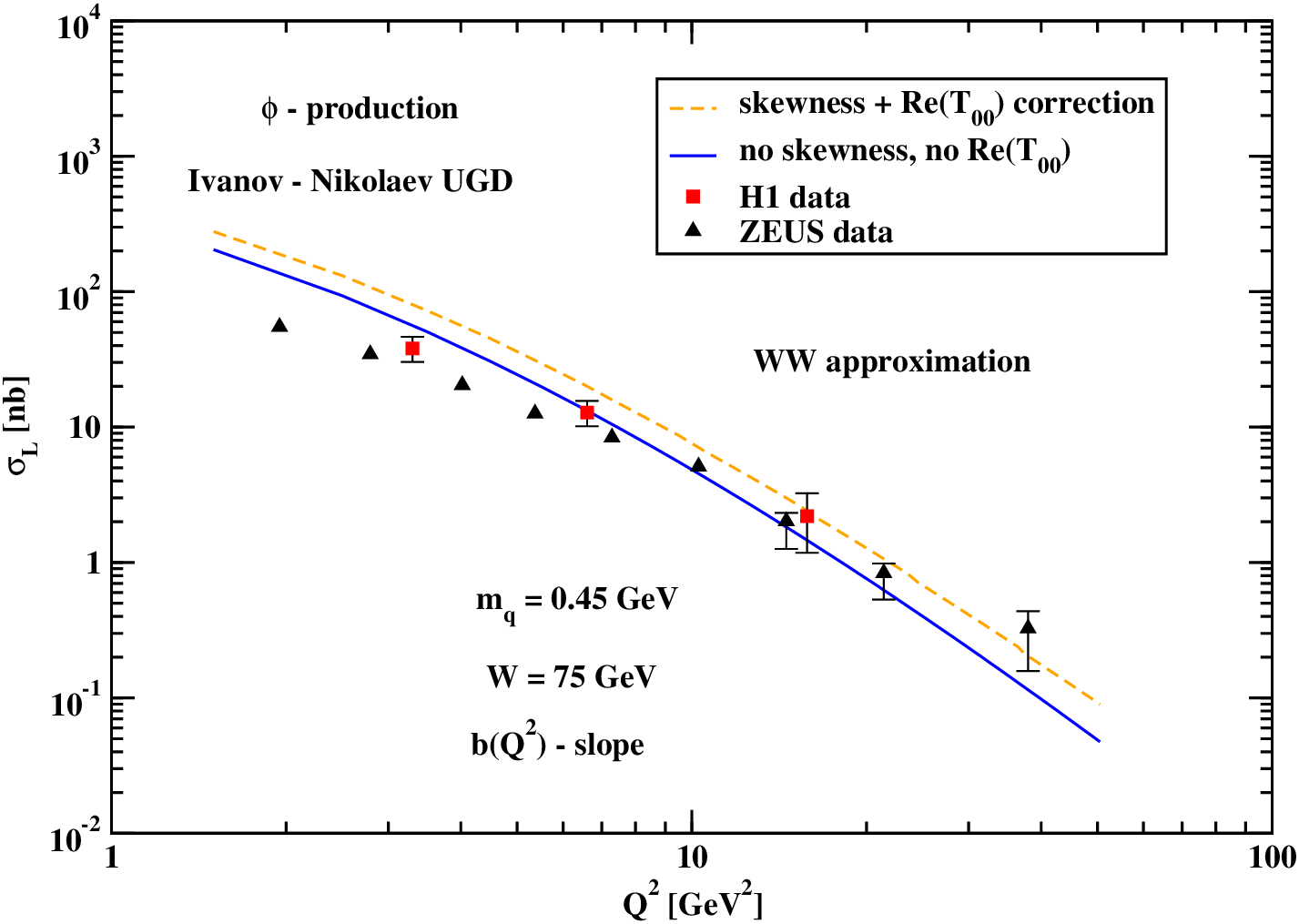}
\includegraphics[scale=0.30,clip]{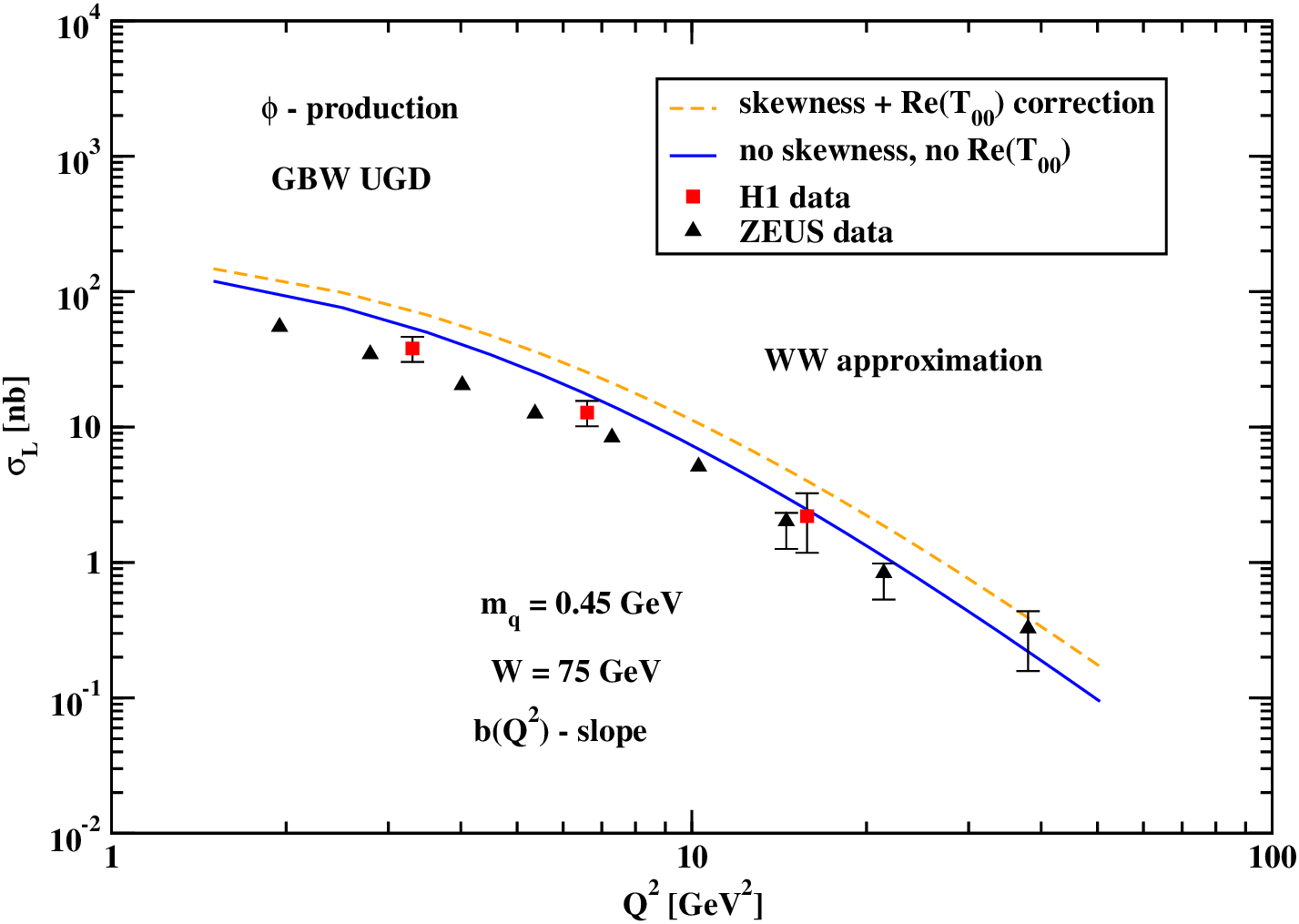}
\caption{An estimation of the skewedness effects and the real part of the amplitude for the longitudinal cross section $\sigma_L$ calculated in the WW approximation using the Ivanov-Nikolaev (left panel) and GBW (right panel) UGDs.}
\label{fig:sigL_skewness}	
\end{figure}

\begin{figure}[htb]
\centering
\includegraphics[scale=0.30,clip]{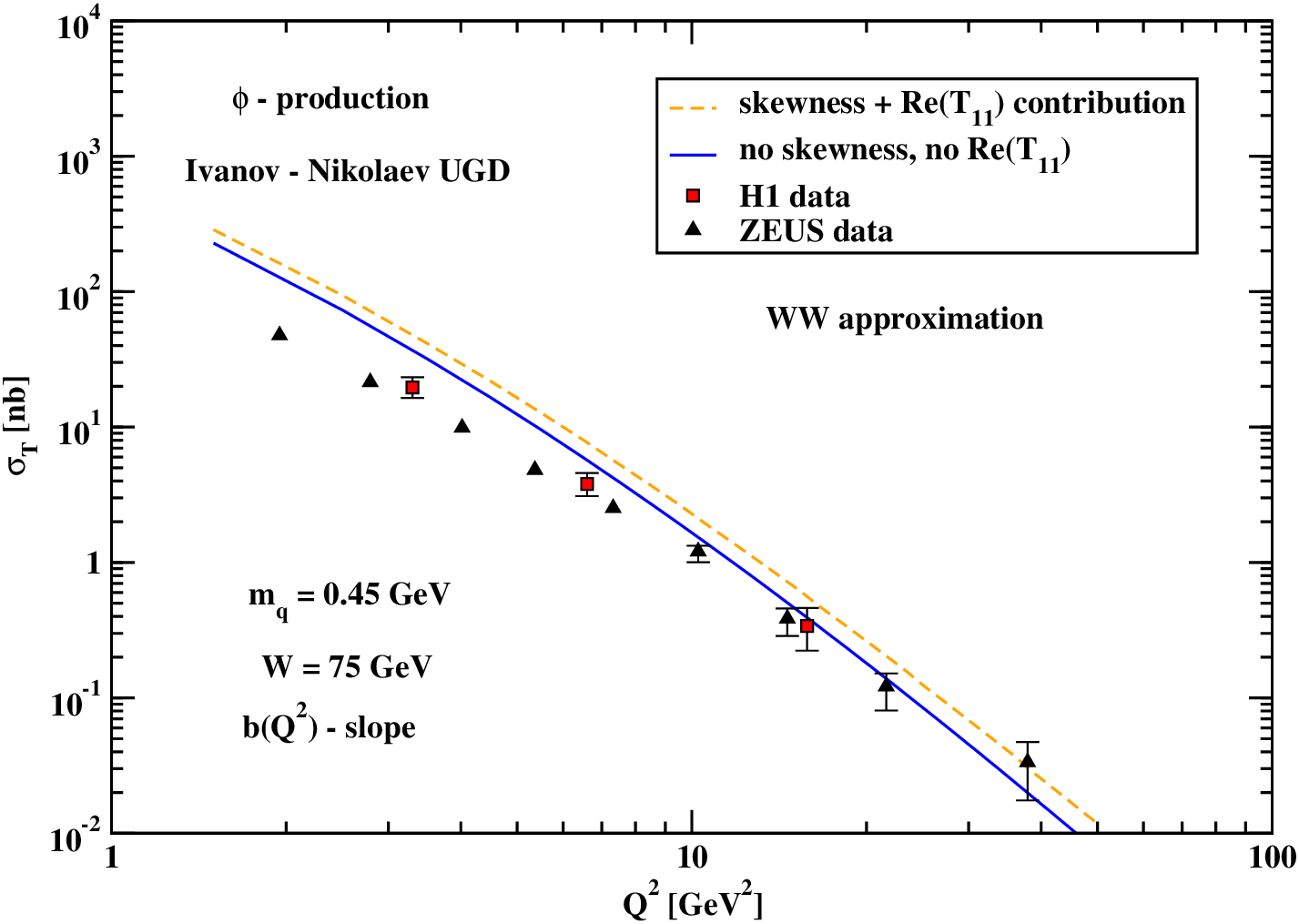}
\includegraphics[scale=0.30,clip]{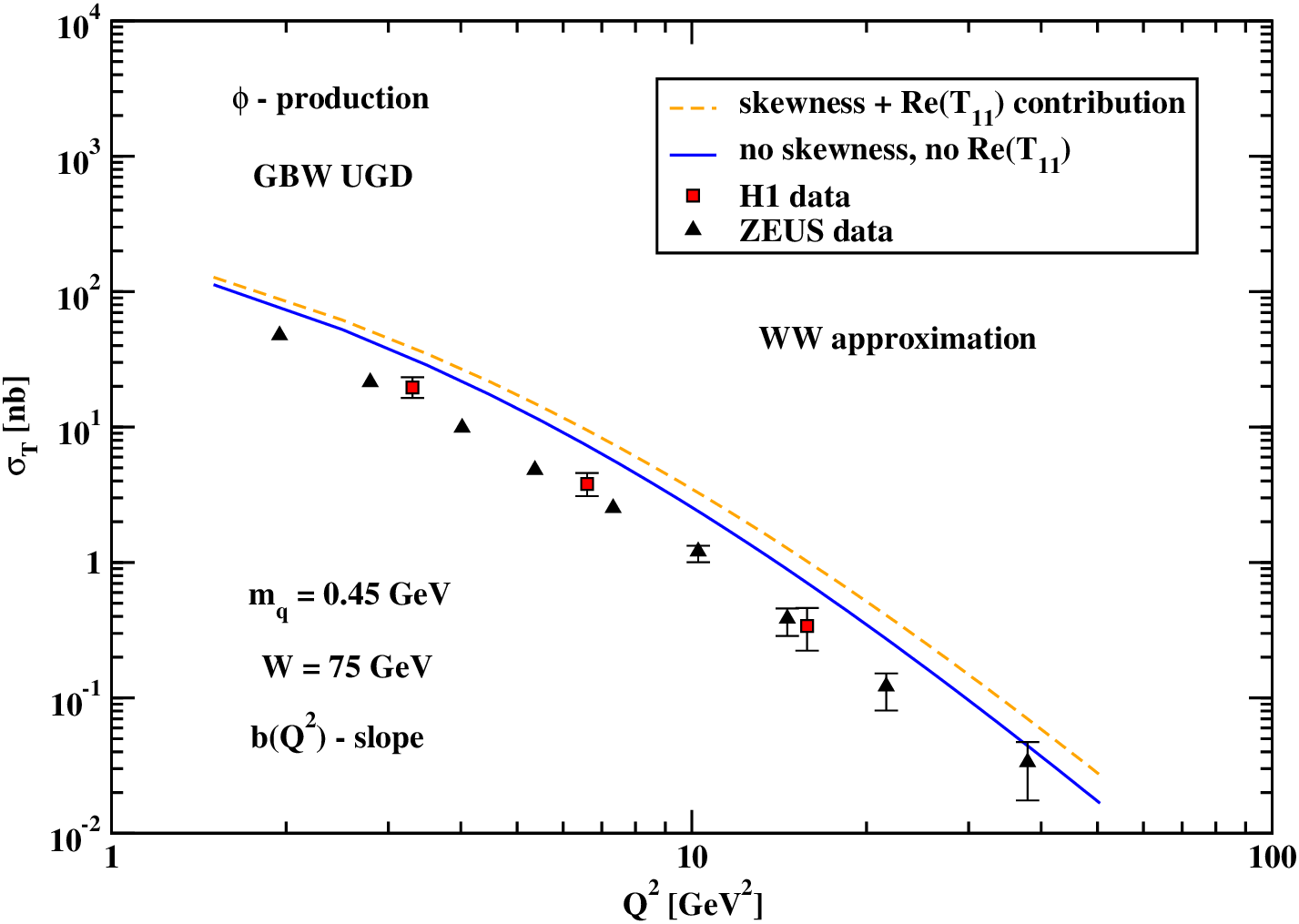}
\caption{An estimation of the skewedness effects and the real part of the amplitude for the transverse cross section $\sigma_T$  calculated in the WW approximation using the Ivanov-Nikolaev (left panel) and GBW (right panel) UGDs.}
\label{fig:sigT_skewness}	
\end{figure}

\begin{figure}[htb]
	\centering
	\includegraphics[scale=0.30,clip]{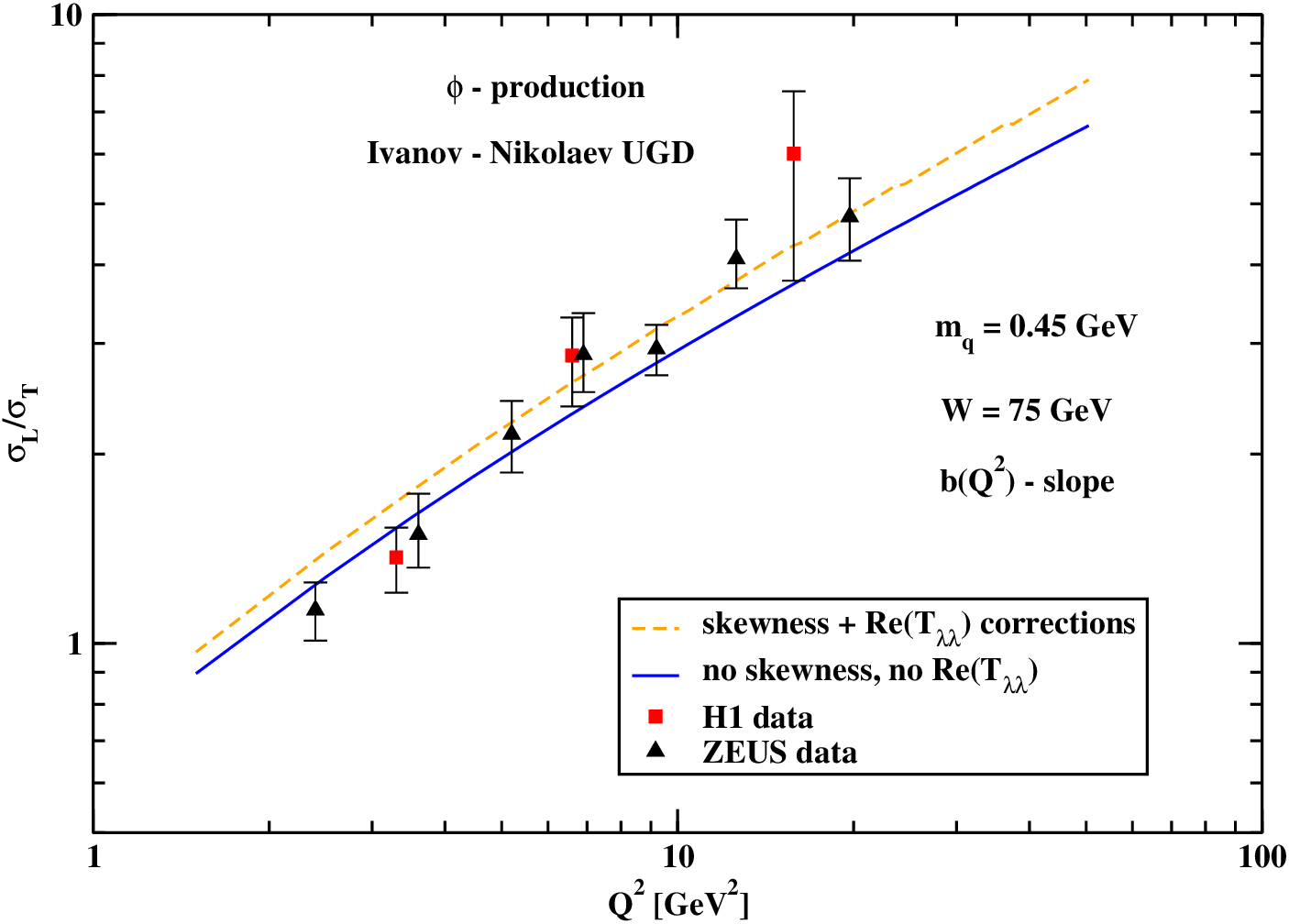}
	\caption{An estimation of the skewedness effects and the real part of the amplitude are shown for the cross sections ratio $\sigma_L/\sigma_T$ through the Ivanov-Nikolaev UGD.}
	\label{fig:ratio_comparison_skewness}	
\end{figure}

So far we did not consider skewedness effects and the real part of the $\gamma^*\,p \to \phi\,p$ amplitude.
Both these corrections can be calculated from the energy dependence of
the forward amplitude. Defining
\begin{eqnarray}
\Delta_\Pom = {\partial \log 
 \Big(	\Im m T_{\lambda_V\lambda_\gamma}(s,Q^2)/s \Big)
 \over \partial \log(1/x)} \, , 
\end{eqnarray}
we can calculate the real part from 
\begin{eqnarray}
\rho = {\Re e  T_{\lambda_V\lambda_\gamma}(s,Q^2) \over \Im m  T_{\lambda_V\lambda_\gamma}(s,Q^2)} = \tan\Big( {\pi \Delta_\Pom \over 2} \Big) \, .
\end{eqnarray}
The skewedness correction is obtained from multiplying the forward amplitude by the factor \cite{Shuvaev:1999ce}:
\begin{eqnarray}
R_{\rm skewed} = { 2^{2 \Delta_\Pom + 3} \over \sqrt{\pi} } \cdot {\Gamma(\Delta_\Pom + 5/2) \over \Gamma(\Delta_\Pom+ 4)} \, .
\end{eqnarray}
Now we wish to show our estimates of these corrections.
We show results for longitudinal (Fig.~\ref{fig:sigL_skewness}) and transverse (Fig.~\ref{fig:sigT_skewness}) components separately. The effect is not too big but cannot
be neglected. The effect of the skewedness is much larger than the effect
of the inclusion of the real part.
We observe that the effect of the skewedness does not cancel in the $\sigma_L/\sigma_T$ ratio as can
be seen in Fig.~\ref{fig:ratio_comparison_skewness}.


	
\newpage

\section{Conclusions} 

In the present paper we have used a recently formulated hybrid 
formalism for the production of $\phi$ meson in the 
$\gamma^*\,p \to \phi\,p$ reaction using unintegrated gluon distributions and meson distribution amplitudes.
In this formalism the $\gamma^*  \to \phi$ impact factor
is calculated in collinear-factorization using (collinear)
distribution amplitudes. So far this formalism was used
only for massless quarks/antiquarks (e.g. for $\rho^0$ meson production).
Both twist-2 and twist-3 contributions are included.
The impact factor for the $p \to p$ transition are expressed in terms
of UGDs. Two different UGD models have been used.

We have shown that for massless quarks the genuine three-parton contribution is more
than order of magnitude smaller than the WW one.
Therefore in this paper we have concentrated on the WW component. 

We have observed too quick rise of the cross section
when going to smaller photon virtualities compared to the experimental
data measured by the H1 and ZEUS collaborations at HERA.
This was attributed to the massless quarks/antiquarks.
We have proposed how to include effective quark masses into the formalism.
Corresponding distribution amplitudes were calculated
and have been used in the present approach.
With effective quark mass $m_q \sim$ 0.5 GeV a good description
of the H1 and ZEUS data has been achieved
for the Ivanov-Nikolaev and GBW UGDs down to $Q^2 \sim$ 4~GeV$^2$.
This value of the strange quark mass is similar as the one found 
in Ref.~\cite{Cisek:2010jk},
where the $k_T$-factorization formalism with $s \bar s$ light-cone wave 
function of the $\phi$ meson was used for real photoproduction. 

We have estimated also the skewedness effect which turned out to be not
too big but not negligible. We have shown some residual effect of
the skewedness for the ratio of longitudinal-to-transverse cross sections. 

\vspace{0.5cm}

{\bf Acknowledgment}
A.D. Bolognino thanks University of Cosenza and INFN for support of her stay in Krak\'ow.
We are indebted to Alessandro Papa for a discussion.\\
This study was partially supported by the Polish National Science Center
grant UMO-2018/31/B/ST2/03537 and by the Center for Innovation and
Transfer of Natural Sciences and Engineering Knowledge in Rzesz{\'o}w.


\end{document}